\documentclass[fleqn,usenatbib]{mnras}

\usepackage{newtxtext,newtxmath}

\usepackage[T1]{fontenc}

\usepackage{graphicx}	
\usepackage{amsmath}	
\usepackage{amssymb}	
\usepackage{natbib}
\usepackage[super]{nth}
\usepackage{wrapfig}

\newcommand\treal{T$_{\rm real}$}
\newcommand\mreal{M$_{\rm real}$}
\newcommand\tfit{T$_{\rm fit}$}
\newcommand\mfit{M$_{\rm fit}$}
\newcommand\bfit{$\beta_{\rm fit}$}
\newcommand\fw{f$_{\rm w}$}
\newcommand\tcmb{T$_{\rm CMB}$}
\defcitealias{Mennella+98}{M98}
\defcitealias{Demyk+17A}{D17A}
\defcitealias{Demyk+17B}{D17B}
\defcitealias{Draine&Li07}{DL07}
\defcitealias{Compiegne+11}{C11}
\defcitealias{Agladze+96}{Agladze et al. 1996}
\defcitealias{Coupeaud+11}{Coupeaud et al. 2011}

\title[Systematic errors in dust fits]{Systematic errors in dust mass determinations: Insights from laboratory opacity measurements}

\author[Fanciullo et al.]{
Lapo Fanciullo,$^{1}$\thanks{Contact e-mail: lfanciullo@asiaa.sinica.edu.tw}
Francisca Kemper,$^{2,1}$
Peter Scicluna,$^{3,1}$
Thavisha E. Dharmawardena$^{4,1,5}$
\newauthor
and Sundar Srinivasan$^{6,1}$
\\
$^{1}$Institute of Astronomy and Astrophysics, Academia Sinica, 11F of AS/NTU Astronomy-Mathematics Building, No.1, Sec. 4, Roosevelt Rd, Taipei 10617, Taiwan, R.O.C.\\
$^{2}$European Southern Observatory, Karl-Schwarzschild-Str.~2, 85748 Garching, Germany\\
$^{3}$European Southern Observatory, Alonso de Cordova 3107, Santiago Regio Metropolitana, Chile\\
$^{4}$Max Planck Institute for Astronomy, K\"onigstuhl 17, 69117 Heidelberg, Germany\\
$^{5}$National Central University, No. 300, Zhongda Rd., Zhongli District, Taoyuan City 32001, Taiwan, R.O.C.\\
$^{6}$Instituto de Radioastronom\'ia y Astrof\'isica, UNAM, Apdo.~Postal 72-3 (Xangari), Morelia, Michoac\'an 58089, Michoac\'{a}n, M\'{e}xico
}

\date{Accepted XXX. Received YYY; in original form ZZZ}

\pubyear{2020}

\begin{document}
\label{firstpage}
\pagerange{\pageref{firstpage}--\pageref{lastpage}}
\maketitle

\begin{abstract}
The thermal emission of dust is one of the most important tracers of the interstellar medium: multi-wavelength photometry in the far-infrared (FIR) and submillimeter (submm) can be fitted with a model, providing estimates of the dust mass. The fit results depend on the assumed value for FIR/submm opacity, which in most models -- due to the scarcity, until recently, of experimental measurements -- is extrapolated from shorter wavelengths.
Lab measurements of dust analogues, however, show that FIR opacities are usually higher than the values used in models and depend on temperature, which suggests that dust mass estimates may be biased. To test the extent of this bias, we create multi-wavelength synthetic photometry for dusty galaxies at different temperatures and redshifts, using experimental results for FIR/submm dust opacity, then we fit the synthetic data using standard dust models. 
We find that the dust masses recovered by typical models are overestimated by a factor 2 to 20, depending on how the experimental opacities are treated. If the experimental dust samples are accurate analogues of interstellar dust, therefore, current dust masses are overestimated by up to a factor of 20. The implications for our understanding of dust, both Galactic and at high redshift, are discussed.
\end{abstract}




\section{Introduction}
\label{section_intro}

Dust is an essential component of the interstellar medium (ISM) despite making up a small fraction of its mass. In addition to playing many roles in the physics and chemistry of the ISM -- such as H$_2$ formation \citep[e.g.][]{Gould&Salpeter63, Wakelam+17}, ice chemistry \citep{Boogert+15}, and gas heating and cooling \citep{WD01A} -- dust is an essential observational tracer. Being well-mixed with interstellar gas, it provides a proxy for the overall gas abundance \citep[e.g.][]{Bohlin+78, Liszt+14}, unlike emission lines such as H{\sc i} or CO, which only trace gas in the atomic or molecular phase, respectively. Building on that, the dust mass is also used as a proxy for star formation rate, which is assumed to  scale with the gas mass. Obtaining dust mass estimates in molecular clouds and on galaxy scales is therefore a very important endeavor in both the local and the high-redshift Universe. This can be seen for instance in the so-called ``dust budget crisis'', where the dust mass estimates are higher than can be comfortably explained by dust formation models (see Sect.~\ref{section_fit_2bands} and  \ref{section_astro_implications}).

In the absence of bright background sources to measure extinction, the only viable way to measure dust masses is to fit a thermal dust emission model to an observed spectral energy distribution (SED). Many dust models exist from which emission can be calculated for given values of dust composition, grain size distribution and intensity of the interstellar radiation field, e.g. \citet{Desert+90, Zubko+04, Draine&Li07, Compiegne+11}; THEMIS \citep{Jones+13, Jones+17}. However, when emission is limited to submillimeter (submm) and far-IR (FIR) wavelengths, it is common to use a simplified model called a modifed blackbody (MBB). In the optically thin limit (which is usually satisfied in the FIR/submm) the flux density for a MBB of temperature T at a distance $D$ follows:
\begin{equation}
\label{eq_mbb_simple}
    F_\nu (\lambda) = \frac{M_d}{D^2} \ \kappa(\lambda) \ B_{\rm \nu}(T) 
\end{equation}
where $M_d$ is the dust mass and $\kappa(\lambda)$ is the wavelength-dependent opacity in the form of a mass absorption coefficient (MAC), i.e. a cross-section per unit mass (such as cm$^2\,$g$^{-1}$). It is very common to express the opacity as a power law: $\kappa(\lambda) = \kappa_0 \ (\lambda/\lambda_0)^{-\beta}$, where $\kappa_0$ is the opacity at $\lambda_0$. The value of $\lambda_0$ can be chosen arbitrarily, although typical choices coincide with the central wavelengths of known FIR/submm broadband filters in the submm like 160, 250, 500 ({\it Herschel}) and $850\,\mu$m ({\it Planck}, SCUBA2). 
The value of the power law index $\beta$ typically falls in the range $1.5 - 2$, but values between 1 and 3 have been reported \citep[e.g.][]{Smith+12, Clements+18}. This type of MBB model with power-law opacity is meant to fit emission from the large and cool dust grain which constitute the bulk of dust mass, and it is used for $\lambda > 50\,\mu$m \citep[see e.g.][]{Casey_12}.

Independently of the type of model used, a fit of dust thermal emission requires that one specifies the dust opacity $\kappa(\lambda)$: since FIR opacity and emission are degenerate, the choice of opacity determines the mass fit result. Despite its importance in SED fits, for a long time FIR/submm opacities have remained poorly understood, largely due to the scarcity of experimental spectra of candidate dust materials in this wavelength range. Consequently, in many dust models the FIR/submm opacity is an extrapolation from shorter wavelengths rather than an experimental quantity. For instance, the $\lambda \gtrsim 100\,\mu$m opacity of the extremely successful \citet{Draine&Li84} graphite and silicate were obtained from the extrapolation of dielectric functions calibrated to reproduce experimental and observational features at $\lambda \lesssim 100\,\mu$m; this remains true after several updates of the dust properties \citep{Li&Draine01, Draine+14}. Over the past few decades, however, the FIR/submm opacity of interstellar dust analogues has been the subject of many laboratory studies, and the experimental results are rather different from the typical opacity used in models: they tend to be higher by up to an order of magnitude, they often do not follow simple power laws and they are dependent on the temperature of the material \citep[e.g.][]{Mennella+98, Coupeaud+11, Demyk+17A, Demyk+17B}. Therefore, there is the possibility that dust masses obtained by SED fits \citep[e.g.][]{Watson+15, Berta+16, Nersesian+19, Aniano+20, DeLooze+20} may be systematically wrong. 

This paper aims to identify and quantify the potential bias on dust masses in the following way: we use experimentally derived $\kappa(\lambda)$ to calculate a synthetic dust SED, we fit the result with a standard method used in observational astronomy, and we compare the parameters of the fit (dust mass and temperature and, where applicable, $\beta$) to the parameters used in the construction of the synthetic SED. The eventual differences between the parameters used in the creation of the SED and those recovered by the fit are an assessment of systematic bias in the dust mass determinations in the nearby and distant universe. The method that we chose to recover dust masses is a MBB fit.

The paper is organized as follows: Section \ref{section_opacity} describes the experimental $\kappa(\lambda)$ selected from the scientific literature for the purpose of this study, as well as our choice of the opacity to use in the MBB fits.
In Section \ref{section_model} we present the synthetic SEDs built from experimental opacities, in the form of FIR/submm photometry for model galaxies with different redshifts and temperature distributions.
Section \ref{section_fit} shows the results of the fits executed on the synthetic photometry and compares the fit parameters to the ones used for the construction.
Section \ref{section_astro_implications} discusses the relevance of the results in an astrophysical context and examines the necessity of coherence with other dust tracers. 
Finally, Section \ref{section_conclusion} summarizes our results and points to future directions.

\section{Material opacity data}
\label{section_opacity}

\subsection{Opacity data selection and main characteristics}
\label{section_opacity_selection}

\begin{table*}
\begin{minipage}{\textwidth}
\caption{Table of materials. The opacity of all materials has been measured at  ${\rm T} = 10,\ 30,\ 100,\ 200$ and 300 K, with the exception of the materials from \citetalias{Mennella+98}, which have been measured at ${\rm T} = 24,\ 100,\ 160,\ 200$ and 295 K.}
\centering
\begin{tabular}{lcccccc}
    Name & Material\footnote{Legend: C = carbon, Sil = silicate, am = amorphous, cr = crystalline.} & Stoichiometry  & $\lambda$ range & References \\ 
    
     &  & (approximate) & $\mu$m &  & \\
    \hline
    
    AC\footnote{Also called ACAR in \citetalias{Mennella+98}.} & C, am & -- & $20 - 2000$ & \citetalias{Mennella+98} \\
    
    BE & C, am & -- & $20 - 2000$ & \citetalias{Mennella+98} \\
    
    FOR & Sil, cr & Mg$_{1.8}$Fe$_{0.2}$SiO$_4$ & $20 - 2000$ & \citetalias{Mennella+98} \\
    
    FAY & Sil, cr & Mg$_{0.12}$Fe$_{1.88}$SiO$_4$ & $20 - 2000$ & \citetalias{Mennella+98} \\
    
    FAYA & Sil, am & Mg$_{0.18}$Fe$_{1.82}$SiO$_4$ & $20 - 2000$ & \citetalias{Mennella+98} \\
    
    X35 & Sil, am & Mg$_{2}$SiO$_4$ & $5 - 1000$ & \citetalias{Demyk+17A} \\
    
    X40 & Sil, am & Mg$_{1.5}$SiO$_3.5$ & $5 - 1000$ & \citetalias{Demyk+17A} \\
    
    X50(A,B)\footnote{Two different MgSiO$_3$ samples, synthesized with different methods, were studied in \citetalias{Demyk+17A} for comparison. Sample X50A was made by melting oxides with a CO$_2$ laser, same as X35 and X40; X50B was made by melting and quenching SiO$_2$ and MgCO$_3$ in a crucible.} & Sil, am & MgSiO$_3$ & $5 - 1000$ & \citetalias{Demyk+17A} \\
    
    E10(R)\footnote{Two different samples were produced in D17B for each stoichiometry. In samples E\textsc{xx} (\textsc{xx} = 10, 20, 30 and 40) ferric iron Fe$^{3+}$ is dominant. Samples E\textsc{xx}R, produced by the reduction of E\textsc{xx}, are richer in ferrous iron Fe$^{2+}$.} & Sil, am & Mg$_{0.9}$Fe$_{0.1}$SiO$_3$ & $5 - 1000$ & \citetalias{Demyk+17B} \\
    
    E20(R) & Sil, am & Mg$_{0.8}$Fe$_{0.2}$SiO$_3$ & $5 - 1000$ & \citetalias{Demyk+17B} \\
    
    E30(R) & Sil, am & Mg$_{0.7}$Fe$_{0.3}$SiO$_3$ & $5 - 1000$ & \citetalias{Demyk+17B} \\
    
    E40(R) & Sil, am & Mg$_{0.6}$Fe$_{0.4}$SiO$_3$ & $5 - 1000$ & \citetalias{Demyk+17B} \\

    \hline
\end{tabular}
\label{tab:materials}
\end{minipage}
\end{table*}

To construct synthetic SEDs we searched the scientific literature for experimentally-measured FIR/submm opacities for a variety of plausible analogues of interstellar dust. We decided on three requirements for these measurements: they must be available at temperatures typical of the cold ISM ($20 - 100\,$K); they must cover the $50\, \mu$m$ - 1\,$mm wavelength range; and the opacity must be available as a MAC rather than a complex refractive index ($n$, $k$). This latest requirement is due to the fact that, to our knowledge, no carbon opacity in ($n$, $k$) form is publicly available available for cryogenic temperatures.\footnote{
\citet{Zubko+96} did derive the (n, k) for three types of amorphous carbon from the experimental MACs of \citet{Colangeli+95}. These (n, k) data, used in the \citet{Zubko+04} and \citet{Compiegne+11} dust models, are however temperature-independent and we assume they have been measured at room temperature. The opacity for two of the \citet{Colangeli+95} carbonaceous materials have been measured at cryogenic temperatures by \citet{Mennella+98}, but only in MAC format.}
The conversion between (n, k) and MAC is not trivial, as it depends on the grains' shape distribution and structure, such as whether they form aggregates \citep[e.g.][see also further discussion in Sect.~\ref{section_model_redmac}]{Boh+Huf83, Stognienko+95}. We therefore chose to only use opacity in MAC format in the present article for self-consistency.

\begin{figure*}
\begin{minipage}{\textwidth}
\begin{center}
\includegraphics[width=0.9\hsize]{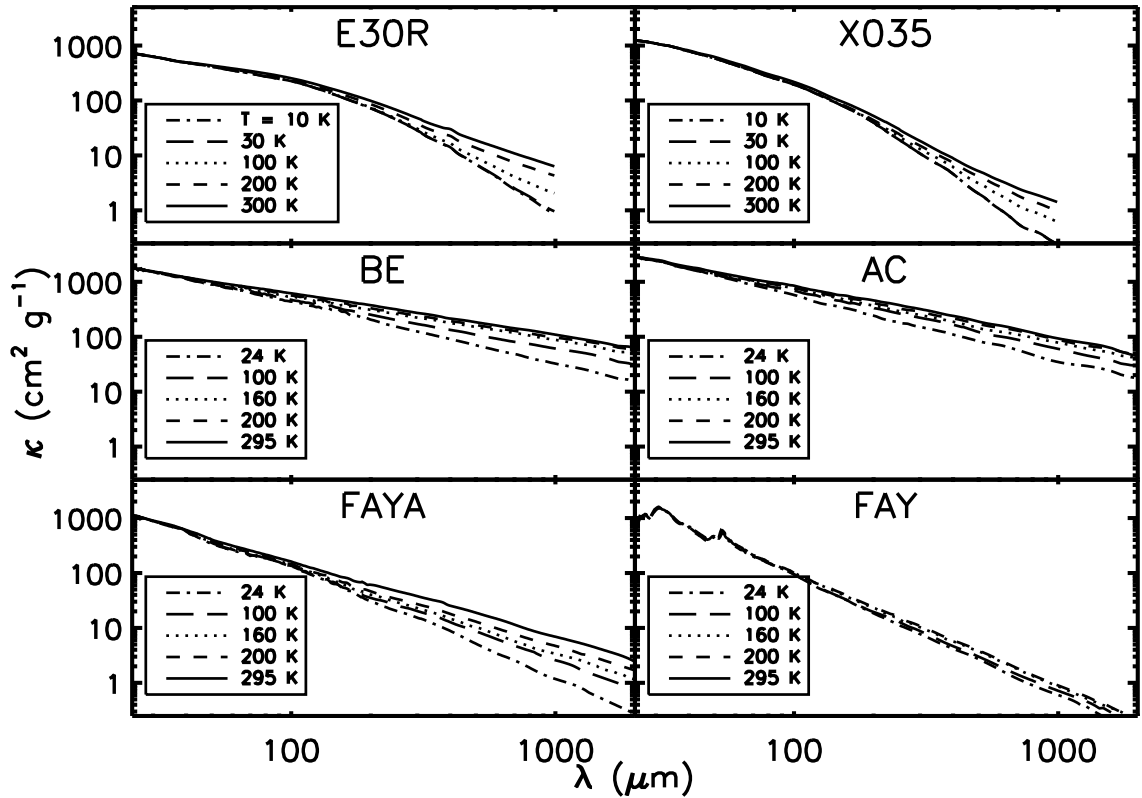}
\caption{
Mass absorption coefficients $\kappa$ for a subsample of the materials from the literature, interpolated to a common wavelength grid. Note that FAYA and FAY have similar chemical compositions but are amorphous and crystalline, respectively (see Table \ref{tab:materials}). 
}
\label{Fig_Opacity_labdata}
\end{center}
\end{minipage}
\end{figure*}

Following these constraints, we made a final selection including the work by \citet{Mennella+98} (hereafter \citetalias{Mennella+98}) 
and \citet{Demyk+17A, Demyk+17B} (hereafter \citetalias{Demyk+17A}, \citetalias{Demyk+17B}). These three studies all show that opacity -- both its absolute value and its dependence on wavelength -- is temperature-dependent. The dust analogue materials from these studies are summarized in Table \ref{tab:materials}. With the exception of the crystalline fayalite (FAY) and crystalline forsterite (FOR), all materials studied are amorphous. Fig.~\ref{Fig_Opacity_labdata} shows the opacity for a subsample of the materials, where the dependence of optical properties on composition and with temperature is evident. 

 The opacity depends not only on the material composition, stoichiometry and microscale structure (e.g. whether the material is amorphous or crystalline), but also on the temperature and -- for Mg+Fe silicates -- the oxidation state of iron. A dependence on the synthesis technique is also observed, which is due to the fact that different techniques may produce different structures on microscopic scales. There is no obvious systematic description for this variation, but a few trends have become evident over the years \citepalias[e.g.][see also Fig.~\ref{Fig_Opacity_labdata}]{Agladze+96, Mennella+98, Coupeaud+11, Demyk+17A, Demyk+17B}:
\begin{itemize}
    \item Opacity is independent of temperature up to $\lambda \sim 30\,\mu$m; for longer wavelengths, the $\kappa$ of amorphous materials is temperature-dependent. This dependence is absent or less pronounced in crystalline materials.
    \item For amorphous materials and for T $>$ 30 K, the opacity increases with temperature.
    \item For T $<$ 30 K the situation is less straightforward. \citet{Coupeaud+11}, \citetalias{Demyk+17A} and \citetalias{Demyk+17B} find no change in opacity between 10 and 30 K; however, \citet{Agladze+96}\footnote{We do not use \citet{Agladze+96} data in the present work because it is only available in the 0.7--2.9 mm wavelength range, which is too long for our purposes.} find that in several materials opacity reaches a minimum at T $\sim$ 20 K and increases again for T $\rightarrow$ 0.
    \item Amorphous materials have higher opacity than crystalline materials of the same stoichiometry.
    \item While silicate opacity changes depending on the iron content and its oxidation state, no systematic trend in this change is apparent.
\end{itemize}

\subsection{Opacity data reprocessing}
\label{section_opacity_reprocessing}

The data as they are, coming from different labs and teams, have been subjected to different types of reduction and can be difficult to compare, so we reprocessed the data to make them more uniform. Note that this reprocessing is for the sake of coding and computational simplicity in the subsequent analysis, and does not affect the validity of the physical results. The final product of the process described in this section is a database of $\kappa(\lambda)$ for different materials and temperatures, regridded on a common wavelength array and with missing data interpolated and smoothed out.

\begin{figure}
\begin{minipage}{0.5\textwidth}
\begin{center}
\includegraphics[width=0.9\hsize]{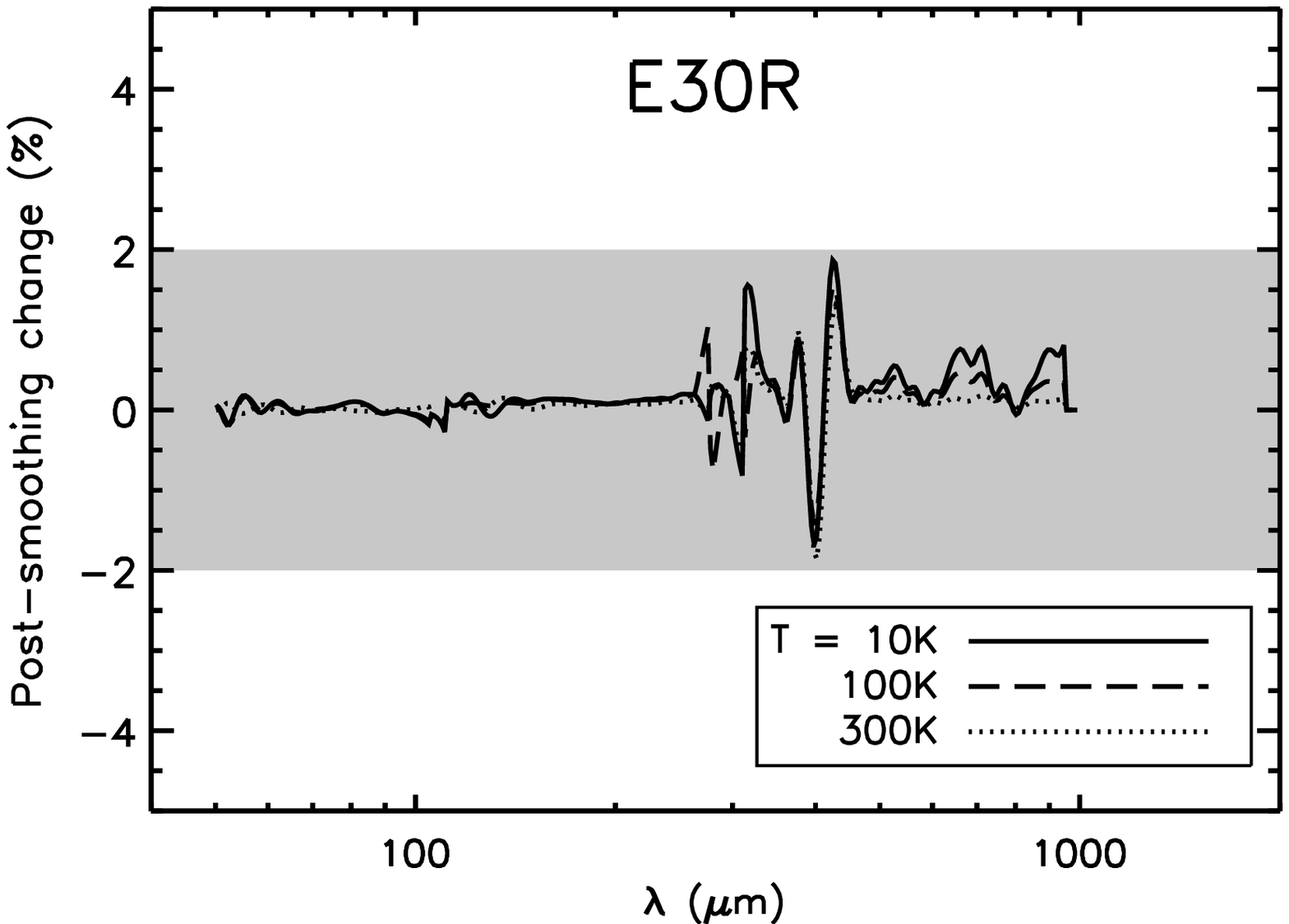}
\includegraphics[width=0.9\hsize]{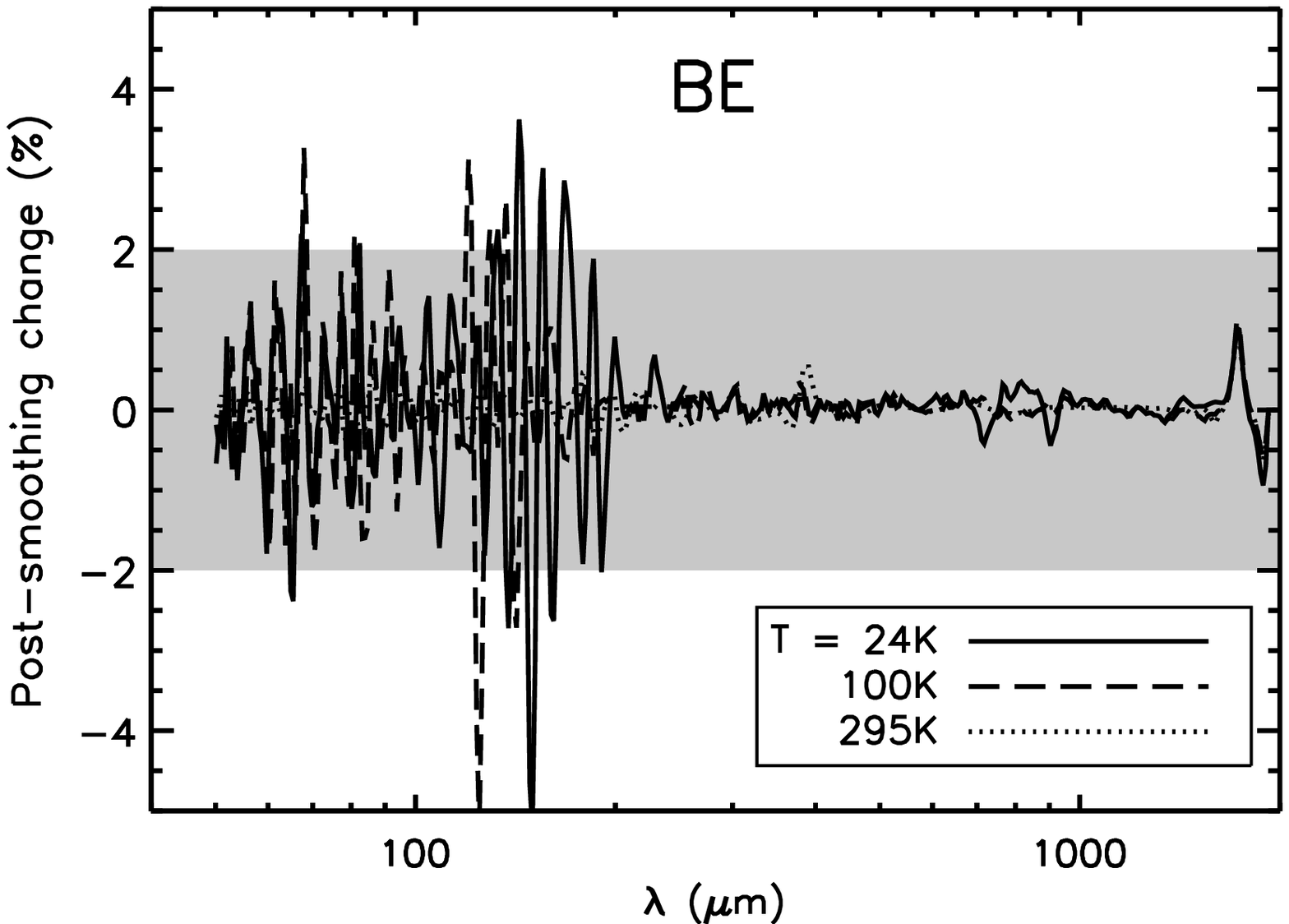}
\caption{
Difference (percentage) between pre-smoothing and post-smoothing opacity, for E30R silicates and BE carbon. The grey shaded area encloses variations of $\pm 2$ per cent.
}
\label{Fig_Smoothing_error}
\end{center}
\end{minipage}
\end{figure}

Some materials have gaps in their wavelength coverage due to the exclusion of lower-quality measurements. To remove these gaps we interpolate $\kappa(\lambda)$ at fixed temperature using a simple power law on wavelength. The sections of missing data cover relatively small wavelength ranges, and as such we do not expect that this introduces a significant source of uncertainty.
Where the missing section is at the beginning or the end of the spectrum, and no interpolation over wavelength is possible, we instead fill the gaps with an interpolation over temperature. Since the relation between temperature and opacity at fixed $\lambda$ is roughly linear, as observed in \citetalias{Mennella+98}, we use a linear interpolation over $T$. This kind of interpolation can introduce a discontinuity into the data, which we remove by smoothing $\kappa(\lambda)$ with a boxcar kernel using a width of 10 steps. We only smooth over wavelength, since the data is much more densely populated in wavelength ($\sim$250 nodes per order of magnitude) than in temperature (3 -- 5 nodes per order of magnitude). The systematics introduced by the smoothing are small; the value of opacity varies by less than 2 per cent in most cases (see Fig.~\ref{Fig_Smoothing_error}). 

In the case of \citetalias{Mennella+98} data the lowest temperature available is 24 K; we make the simplifying assumption that \citetalias{Mennella+98} opacities are constant for ${\rm T} < 24\,$K, which is consistent with the \citet{Coupeaud+11}, \citetalias{Demyk+17A} and \citetalias{Demyk+17B}  observations that opacities are constant for $T < 30\,$K. We will not use dust temperatures lower than 20 K in this paper; thus, even if we expect that material opacities reach a minimum at $T \sim 20\,K$ as per \citet{Agladze+96}, the error introduced by the extrapolation is small, since it takes place near the minimum of a function.

It should be noted that, since $\kappa$ depends on the shape and structure of the grains, using the values directly measured in the lab for our work -- as we do in Sect.~\ref{section_model} and \ref{section_fit} -- implicitly assumes that interstellar grains have the same structure as the dust analogues, i.e. irregularly-shaped aggregates (see e.g. \citetalias{Demyk+17B}). This is a good approximation in cloud cores \citep[e.g.][]{Ormel+09, Koehler+15} but not necessarily in the diffuse ISM; Sect.~\ref{section_model_redmac} shows how we take this complication into account in our model. 

\subsection{Comparison with standard dust models}
\label{section_opacity_compare}

\begin{figure}
\begin{minipage}{0.5\textwidth}
\begin{center}
\includegraphics[width=.99\hsize]{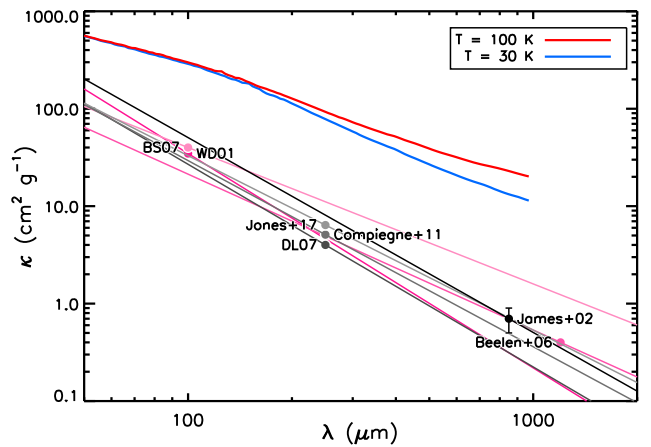}
\caption{
Laboratory-derived mass absorption coefficient (red and blue curves) for the standard composition used in the present article. Pink, grey and black lines are $\kappa$ values from the literature shown for comparison, with circles indicating the ($\lambda_0$, $\kappa_0$) values. Some of the reference opacities are power-law fits to the original (see text). {\it References:} WD01 = \citet{WD01B}; James+02 = \citet[][, with error bars on $\kappa_0$]{James+02}; Beelen+06 = \citet{Beelen+06}; BS07 = \citet{Bianchi&Schneider07}; \citetalias{Draine&Li07} = \citet{Draine&Li07}; Compiegne+11 = \citet{Compiegne+11}; Jones+17 = THEMIS \citep{Jones+17}. 
}
\label{Fig_Opacity_labvslit}
\end{center}
\end{minipage}
\end{figure}

Fig.~\ref{Fig_Opacity_labvslit} compares the experimental opacity for our standard model (70 per cent E30R silicate + 30 per cent BE carbon; see Sect.~\ref{section_model}) to a representative set of $\kappa$ from published interstellar dust models. We feature experimental opacity at two different temperatures -- 30 and 100 K -- to show its dependence on temperature. The comparison opacities include three models calibrated on high-latitude Milky Way dust: \citet{Draine&Li07}, \citet{Compiegne+11} and THEMIS \citet{Jones+13, Jones+17}; three models based on high-redshift sources or analogues: \citet[][the Small Magellanic Cloud]{WD01B}, \citet[][QSOs]{Beelen+06}, and \citet[][supernova models]{Valiante+11}; and one model calibrated on nearby galaxies \citet{James+02}. In the case of the Milky Way and Small Magellanic Cloud models, while the opacity does not strictly follow a power law, it is still well approximated by one; Fig.~\ref{Fig_Opacity_labvslit} therefore shows a power-law fit to the opacity of \citet{WD01B} \citep[from][]{Bianchi&Schneider07}, \citet{Draine&Li07} and \citet{Compiegne+11} \citep[from][]{Bianchi_13}, and THEMIS \citep[from][]{Galliano+18}.
As can be seen from Fig.~\ref{Fig_Opacity_labvslit}, the opacity we obtained from experimental data is much higher than any we found in typical dust models; we expect that any SED fit using the lab-derived opacity will therefore give much different dust mass estimates. 

In the rest of the paper, we use the opacity from \citet{James+02} -- $\kappa_0 = 0.7 \pm 0.2\,{\rm cm}^2\,{\rm g}^{-1}$ at $\lambda_0 = 850\,\mu$m, obtained by calibrating dust opacity against $850{\rm -}\mu$m emission and estimated dust masses from elemental depletion in 22 galaxies -- as representative of typical value from the scientific literature. This does not necessarily mean that the \citet{James+02} determination is the most accurate among those presented; however, since it falls in the middle of the range for the $\kappa$ values at $850 \mu$m, it is a good choice for a representative value of opacities from the scientific literature.

\section{Model: synthetic photometry}
\label{section_model}

The purpose of our project is to construct dust emission photometry for synthetic sources spanning a variety of temperature distribution and redshifts, then fit said photometry with a representative dust model from the astrophysical literature to test how the results depend on the above factors. We model our sources as pointlike, to represent unresolved galaxies, with dust masses of $10^8\,{\rm M}_\odot$. We use redshift values ranging between 0 and 7 and dust temperatures between 20 and $100\,$K. We treat our galaxies as optically thin in the FIR/submm \citep[following][]{daCunha+13} and leave the optically thick case for a follow-up. The redshift effects are calculated assuming a $\Lambda$CDM cosmology with ${\rm H}_0 = 70 {\rm km}\,{\rm s}^{-1}\,{\rm Mpc}^{-1}$, $\Omega_{\rm M} = 0.3$ and a cosmic microwave background (CMB) temperature T$_{\rm CMB} = 2.725\,$K at z~=~0.  

We compute our synthetic SEDs using a generalization of Eq.~\ref{eq_mbb_simple} to multiple materials and temperatures; for $\kappa(\lambda)$ we use the smoothed lab opacities described in Sect.~\ref{section_opacity}. For a redshift z we then have:
\begin{equation}
\label{eq_mbb_multiple}
    F_\nu (\lambda) = \frac{1+z}{D_L^2}\, M_d \sum_{i=1}^{n_{T}} \xi_i \ B_{\rm \nu}\left(\frac{\lambda}{1+z}, T_{i} \right) \sum_{j=1}^{n_{mat}} \eta_j \ \kappa_j\left( \frac{\lambda}{1+z}, T_i \right)
\end{equation}
where $\lambda$ is the wavelength in the reference frame of the observer. In this equation, the dust is composed of $n_{mat}$ different species, each with its own temperature-dependent opacity $\kappa_j$ and weighting fraction $\eta_j$; the temperature distribution is discrete and each of the $n_{T}$ temperatures has a weighting fraction $\xi_i$. For simplicity, in the present work we use the same temperature distribution for all dust species.\footnote{Note that the way Eq.~\ref{eq_mbb_multiple} is written -- the sum on materials being inside the sum on temperatures -- imposes that all materials have the same $T_i$ and $\xi_i$.} The distance factor is $(1+z)/D_L^2$, where $D_L$ is the luminosity distance \citep{Peacock_99}. We arbitrarily set $D_L = $ 100 Mpc for z = 0. At high z, Eq.~\ref{eq_mbb_multiple} needs an additional wavelength-dependent correction to account for the CMB background subtraction, as explained in Sect.~\ref{section_model_CMB}.

Following typical Milky Way dust models \citep[e.g][]{WD01B, Compiegne+11} we decided to make our model 70 per cent silicates and 30 per cent carbon in mass. While there is no assurance that this composition would be realistic for the early Universe, the limited knowledge available about high-redshift dust means that high-z galaxy SEDs are often fit using models calibrated on the local Universe \citep[e.g.][]{Berta+16, Magdis+17}. In the present work we use the amorphous carbon BE from \citetalias{Mennella+98} and the amorphous silicate E30R (partly reduced Mg$_{0.7}$Fe$_{0.3}$SiO$_3$) from \citetalias{Demyk+17B}. The choice of a Mg-Fe silicate is motivated by elemental abundances as measured from depletion, which suggest that the atomic fraction of Mg in dust is similar or slightly larger than that of Fe \citep[e.g.][and refs. therein]{Compiegne+11, Jones+13}. Beyond this, the choice is somewhat arbitrary. We also tested different material compositions for comparison, keeping the same silicate/carbon ratio, and found that the results do not vary significantly.

After creating the synthetic SEDs we turn them into multi-band photometry through convolution with FIR/submm filter profiles. The filters available in our online scripts are the following:
\begin{itemize}
    \item \emph{Herschel} PACS blue ($\lambda = 70\,\mu$m), green ($100\,\mu$m) and red ($160\,\mu$m); 
    \item \emph{Herschel} SPIRE 250, 350 and $500\,\mu$m;
    \item SCUBA2 450 and $850\,\mu$m bands;
    \item ALMA bands 3 to 10, using the default setting for continuum observations in the ALMA Observing Tool\footnote{\url{https://almascience.eso.org/tools/proposing/observing-tool}} for Cycle 6.2 (see Table \ref{tab:filters} for the values of these default settings).
\end{itemize}

In the present work we only show the results for \emph{Herschel} and ALMA bands, which overall cover the wavelength range between $70\,\mu$m and 3.1 mm (see Table \ref{tab:filters}), although each SED only uses a fraction of the bands. The band selection process for each SED is explained in Sect. \ref{section_model_uncertainties}.

\begin{table}
\centering
\begin{minipage}{0.5\textwidth}
\caption{Table of filters used.}
\begin{tabular}{lcc}
Band & Central $\lambda$ & Adopted confusion limit\footnote{Source for {\it Herschel} bands: {\it Herschel} Observers' Manual, Sect. 4.3: \url{http://herschel.esac.esa.int/Docs/Herschel/html/Observatory.html}. Source for SCUBA2 bands: \citet{Dempsey+13}.} \\
  & ($\mu$m) & (mJy/Beam) \\
    \hline
    PACS70 & 70.00 & 0.1 \\
    PACS100 & 100.00 & 0.3 \\
    PACS160 & 160.00 & 1.0 \\
    SPIRE250 & 250.00 & 6.0 \\
    SPIRE350 & 350.00 & 6.0 \\
    SPIRE500 & 500.00 & 7.0 \\
    SCUBA2\_450\footnote{SCUBA2 filters are not used in this article but are available in the online scripts.} & 450.00 & 0.5 \\
    SCUBA2\_850 & 850.00 & 0.7 \\
    ALMA 10 & 344.89 & -- \\
    ALMA 9 & 441.52 & -- \\
    ALMA 8 & 743.90 & -- \\
    ALMA 7 & 872.76 & -- \\
    ALMA 6 & 1286.66 & -- \\
    ALMA 5 & 1476.81 & -- \\
    ALMA 4 & 2067.53 & -- \\
    ALMA 3 & 3074.79 & -- \\
    \hline
\end{tabular}
\label{tab:filters}
\end{minipage}
\end{table}

\subsection{Grain structure effects: raw vs. reduced opacity}
\label{section_model_redmac}

\begin{figure}
\begin{minipage}{0.5\textwidth}
\begin{center}
\includegraphics[width=.99\hsize]{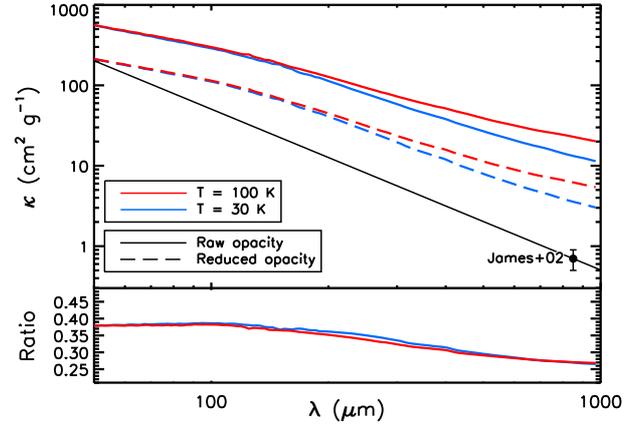}
\caption{
Opacity correction effects (compare Fig.~\ref{Fig_Opacity_labvslit}). {\it Top:} pre- and post-reduction experimental $\kappa$ (red and blue), compared with the \citet{James+02} value (black). {\it Bottom:} ratio of post-reduction to pre-reduction $\kappa$.}
\label{Fig_opcorr}
\end{center}
\end{minipage}
\end{figure}

Using an experimentally-derived MAC as the opacity implies that the dust in our model has the same properties as the samples studied in a laboratory setting. To what point this implicit assumption is valid will be examined in this section. Lab samples typically consist of irregularly-shaped (``fluffy'') aggregates of many particles, which have a higher FIR opacity per unit mass than the individual grains would have \citep[e.g.,][]{Ormel+11, Koehler+15, Ysard+18}. Most interstellar dust models -- especially those calibrated on high-latitude Galactic emission -- assume separate, and usually spherical, dust grains. Therefore, at least part of the difference between the $\kappa$ of experimental studies and that of typical models comes from the assumed grain structure. On the other hand, experimental opacity is not influenced by grain size in our case: dust in the \citetalias{Mennella+98}, \citetalias{Demyk+17A} and \citetalias{Demyk+17B} experimental data is in the Rayleigh regime (particle much smaller than the wavelength) where grains size has no effect on opacity per unit volume. Therefore, we only need correct for aggregates. 

It is debatable whether single-grain opacity or aggregate opacity are best suited to modelling full galaxies, as in the present paper. Many dust models are developed to fit high-latitude dust emission from the diffuse ISM, and assume individual grains. On the other hand, in dense environment such as the interior of dark clouds grains are expected to form aggregates \citep{Stepnik+03,Ysard+13}, and the dense ISM represent a large fraction of the Galactic ISM mass \citep[e.g.][]{Draine_book}.

Nonetheless, it will be informative to repeat our analysis with a ``reduced'' version of our MACs to correct for the effect of aggregates and see what the \mfit/\mreal\ ratio (i.e. the ratio between the mass fit result \mfit\ and the ``real'' mass \mreal\ used in the SED production) would be in a hypothetical aggregate-poor galaxy. We follow the modelling of aggregate opacity from \citet{Ysard+18}, which has the advantage of not being strongly dependent of the characteristics of the aggregate. As shown in \citet{Ysard+18}, Fig.~7 and 8, the far-infrared MAC -- or $\kappa$ -- of aggregates initially increases with the number of grains involved, but quickly plateaus at a value $\sim 2 \times$ that the one of individual grains for silicates, and $\sim 3-4 \times$, with a shallow dependence on wavelength, for carbon (specifically aromatic amorphous carbon or a-C). We create a second set of ``reduced opacity'' synthetic photometry where we divide the silicate MAC by 2 and the carbon MAC by 4, so that the fit results will provide a conservative dust mass estimate. As shown in Fig.~\ref{Fig_opcorr}, the thus-reduced opacity is still $\sim5 \times$ higher than the \citet{James+02} value at $850\mu$m (as opposed to the $\sim 19 \times$ value for unmodified opacity). Fig.~\ref{Fig_opcorr} also shows that, since the opacities of the two dust components (E30R and BE) are reduced by different factors, the overall change in $\kappa$ is not a simple rescaling, but it is wavelength-dependent. In the rest of the article we will refer to the modified $\kappa$ as ``reduced opacity'', and to the unmodified $\kappa$ as ``raw opacity''.

\subsection{Single-temperature vs. two-temperature dust}
\label{section_model_multiT}

The simplest model we can make is one where all dust is at the same temperature. Such a model is appropriate for $\lambda \gtrsim 100\,\mu$m and for typical ``cold dust'' temperatures, but a significant contribution from warm dust -- e.g. stochastically-heated small grains -- is expected at shorter wavelengths \citep[see e.g.][]{Compiegne+11, Jones+13}. We could simply limit our analysis $\lambda \geq 100\,\mu$m, but for warmer dust ($T \gtrsim 40$ K) the SED emission peak would fall outside this wavelength range, resulting in badly constrained fit parameters. To extend our fit to $\lambda < 100\,\mu$m we need to add a second, warm dust component. Therefore, in addition to the synthetic photometry for single-temperature dust in the $100\,\mu{\rm m} < \lambda < 1000\,\mu$m range, we also created photometry for two-temperature dust in the $50\,\mu{\rm m} < \lambda < 1000\,\mu$m range, consisting of a cold component with a temperature of 30 K and a warm component with a temperature of 100 K. We vary the warm mass fraction \fw\ between $10^{-4}$ and 0.3.
Note that the qualifiers "single-temperature" and "two-temperature" refer to the models used to create the synthetic photometry and {\it not} to the model used to fit it: both photometric sets are fit using a single-temperature MBB (see Sect.~\ref{section_fit}). 

\subsection{The effect of the Cosmic Microwave Background}
\label{section_model_CMB}

One of the main distinctions between dust studies at low and high redshift is the effect of the CMB. In the local Universe the temperature of interstellar dust is high enough ($\sim 20\,$K) that thermal dust emission is easily detectable against the CMB. At high redshift, where \tcmb\ increases proportionally to 1+z, the CMB can reach temperatures comparable to those of dust in the local Universe (e.g., T$_{\rm CMB} = 16.35\,$K at z = 5). This has two main consequences \citep{daCunha+13}:
\begin{itemize}
    \item Because of heating by the CMB, dust at high z is warmer than its equivalent  in the local Universe \citep[i.e. dust receiving the same amount of heat from starlight; see Fig.~1 in][]{daCunha+13};
    \item Since dust emission is always measured against the CMB, the SED from high-redshift sources can significantly decrease after background subtraction.\footnote{This effect may be counterintuitive, since the \tcmb\ observed from Earth is always $2.725\,$K. An alternative way of visualizing this is that, e.g., $30\,$K dust at z = 5 has apparent temperature of 5 K when observed from Earth, due to the redshift, and is therefore hard to observe against a $2.725\,$K background. In the limit case where CMB heating is dominant, dust has the same temperature as the CMB and it is therefore undetectable against it.} This effect is stronger at longer wavelengths, so the post-subtraction SED is bluer than the intrinsic one. 
\end{itemize}

In the present work we can ignore the first effect, since we set the dust temperature rather than calculating it from an interstellar radiation field. We decided to only exclude the unphysical situation where T$_{\rm dust} < {\rm T_{CMB}}$, as is the case for T = $20\,K$ at z = 7. As we will see in Sect.~\ref{section_model_uncertainties}, our band selection criteria ensure tighter constraints than that.

The effect of background subtraction needs to be properly implemented in our model, which is done by multiplying the flux from Eq.~\ref{eq_mbb_multiple} by a wavelength-dependent corrective factor. For single-temperature dust and in the optically thin case \citep{daCunha+13} this factor is
\begin{equation}
\label{eq_CMBcorr_singleT}
\frac{F_\nu^{\rm observed}}{F_\nu^{\rm intrinsic}} = 1 - \frac{B_{\rm \nu}[{\rm T_{CMB}(z)}]}{B_{\rm \nu}[{\rm T_{dust}(z)}]}
\end{equation}
From Eqs.~17 and 18 of \citet{daCunha+13} a multi-temperature generalization can be recovered:
\begin{equation}
\label{eq_CMBcorr}
    C_{\rm CMB} = \frac{F_\nu^{\rm observed}}{F_\nu^{\rm intrinsic}} = 1 - \frac{B_{\rm \nu}[{\rm T_{CMB}(z)}]}{\sum_{i=1}^{n_{T}} \xi_{\rm i}\, B_{\rm \nu}[{\rm T_{dust,i}(z)}]}
\end{equation}
where $n_{T}$ is the number of dust temperatures included and $\xi_{\rm i}$ is the correspondent mass fraction (note that the corrective factor is always smaller than unity). Our final synthetic SEDs are therefore the product of Eq.~\ref{eq_mbb_multiple} and Eq.~\ref{eq_CMBcorr}.

\subsection{Model uncertainties and band selection}
\label{section_model_uncertainties}

\subsubsection{SED uncertainties}
\label{section_SED_uncertainties}

The source of uncertainty in our model emission is the uncertainty on the experimental $\kappa$ themselves. This uncertainty includes a statistical and a calibration component. The opacities from \citetalias{Demyk+17A} and \citetalias{Demyk+17B} have a statistical (relative) uncertainty of $0.03\, {\rm cm}^2{\rm g}^{-1}$ and a calibration (absolute) uncertainty of 10 per cent or better. Since our final product will be broad-band (synthetic) photometry, we decided to neglect the statistical uncertainty which is smoothed out by integration over the photometric band, and only keep the calibration component. The uncertainty for \citetalias{Mennella+98} opacities is not mentioned in the article, but Fig.~9 on the same paper shows error bars of $\sim 10$ per cent at $\lambda = 2\,$mm. Since the uncertainty on opacity tends to increase with wavelength, we adopted a systematic uncertainty of 10 per cent on the \citetalias{Mennella+98} values as a conservative estimate. 

An additional uncertainty on the lab-measured opacity is the change on the value of opacity after smoothing (Sect.~\ref{section_opacity_reprocessing}). This change is usually less than $\pm 2$ per cent and never larger than $\sim 5$ per cent and it is narrow-band in nature, so it can be neglected without significant effects on the present study. 

A final source of uncertainty on the opacity comes from the interpolation on temperature (Sect.~\ref{section_opacity_reprocessing}). Unfortunately, this specific uncertainty is extremely hard to estimate, since the physical mechanism of temperature-driven variations in dust opacity is not yet fully understood (See e.g.\ \citetalias{Demyk+17B}, Sect.~4.2). We decided to neglect this source of uncertainty in the present work and we caution readers that the error bars for our models may be underestimated when the models temperatures are far from the temperature measured in the lab (see Table \ref{tab:materials}).

\subsubsection{Creating synthetic observations}
\label{section_mock_observations}

Since our aim is to create synthetic photometry for our objects as they would be observed with \emph{Herschel}, SCUBA2 and ALMA, we need to account for the effect of instrumental limitations. 

The precision of ALMA observations is primarily determined by exposure time. Since typical proposals aim for a 5-10 per cent uncertainties, we decided to adopt a 10 per cent uncertainty on the flux on these bands. The noise for {\it Herschel} and SCUBA2 is more complex: due to their larger beams, their noise cannot descend below a minimum level determined by the unresolved background sources (the so-called confusion noise; see Table \ref{tab:filters}). Since our observations are supposed to be point sources, we adopt the noise on one beam as the overall confusion noise. Yet another source of uncertainty is calibration. For {\it Herschel}, this amounts to  5 per cent on PACS bands \citep{Poglitsch+10} and 5.5 per cent on SPIRE bands \citep[SPIRE Handbook v3.1,][Sect.~7.1]{SPIRE_manual}; for SCUBA2, the value is 12 per cent at $450~\mu$m and 8 per cent at $850~\mu$m \citep{Dempsey+13}.

We calculate the overall uncertainty on the synthetic fluxes with a Monte-Carlo run of $10\,000$ cases. In each iteration, the deviations due to systematic uncertainties (material opacity uncertainty, instrumental calibration) are identical on all relevant bands, while the deviations due to stochastic uncertainties (photon noise, confusion noise) are independent from one band to another. For each photometric band we selected the \nth{16}, \nth{50} and \nth{84} percentiles\footnote{Corresponding to the center$-1\sigma$, center, and center$+1\sigma$ of the distribution if it were a Gaussian.} of the  $10\,000$ iterations and we took the \nth{50} percentile, i.e. the median, as the value of the observed flux. It was our intention to take the difference between \nth{50} and \nth{16} percentile (\nth{84} and \nth{50} percentile respectively) as the one-sigma negative (positive) error bar, but after noticing that the error bars were close to symmetric around the median, we decided to use the half-difference between \nth{84} and \nth{16} percentile as the (symmetric) photometric uncertainty.

\subsubsection{Band selection}
\label{section_band_selection}

The signal-to-noise ratio (S/N) for the synthetic fluxes varies significantly across bands; at high redshift and low temperatures, especially, there are several ``non-detections'' (bands with a S/N lower than 3). It was therefore necessary to select the bands to be used in the fit. 
We kept those bands where the S/N is at least 3 and where the central wavelength does not fall below a rest value of $100\,\mu$m ($50\,\mu$m for two-temperature models; see Sect.~\ref{section_model_multiT}), since our model is not realistic at wavelengths shorter than that threshold value. We only fit those models that have four or more bands left after selection; thus we discarded the models with ${\rm T} = 20\,{\rm K}$ at ${\rm z} \geq 6$ and the model with ${\rm T} = 25\,{\rm K}$ at z = 7. The final fit-ready models have between 4 and 8 bands each.

\section{Fit results and discussion}
\label{section_fit}

\begin{figure*}
\begin{minipage}{\textwidth}
\begin{center}
\includegraphics[width=.9\hsize]{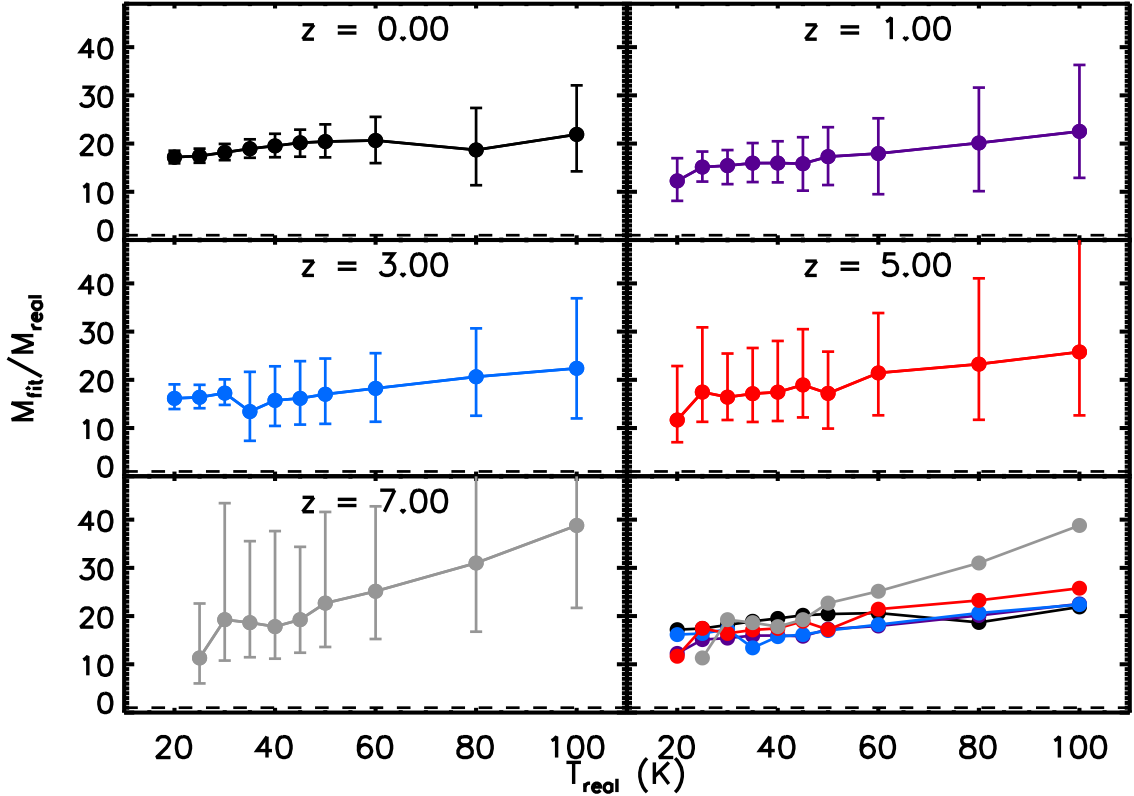}
\caption{
Mass fit results for our dust model (70 per cent E30R silicate, 30 per cent BE carbon) as a function for $z$, for $T_{\rm real}$ between 20 and 100 K. The last panel shows a superposition of all previous plots (minus error bars) for comparison purposes. The equality line \mfit/\mreal\ = 1 is shown as a dashed line. The error bars do not include the 30 per cent uncertainty due to the that comes from the \citet{James+02} $\kappa_0$ itself (see text). Said uncertainty affects all values in the same way, and therefore does not change the trends shown in the figure.
}
\label{Fig_massfit_rawMAC}
\end{center}
\end{minipage}
\end{figure*}

\begin{figure*}
\begin{minipage}{\textwidth}
\begin{center}
\includegraphics[width=\hsize]{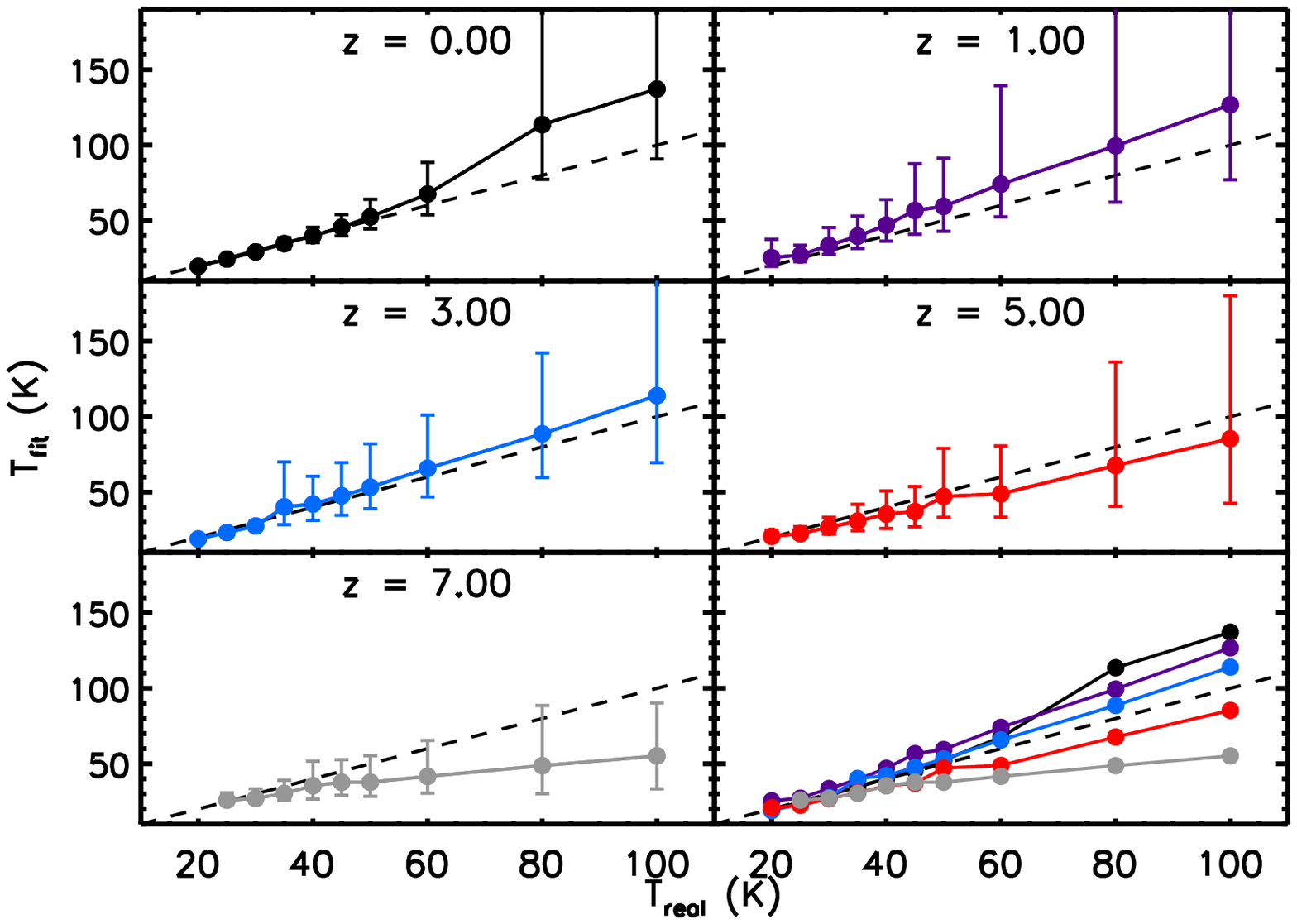}
\caption{
Same as Fig.~\ref{Fig_massfit_rawMAC}, but showing \tfit\ rather than \mfit.
}
\label{Fig_Tfit_rawMAC}
\end{center}
\end{minipage}
\end{figure*}

\begin{figure*}
\begin{minipage}{\textwidth}
\begin{center}
\includegraphics[width=\hsize]{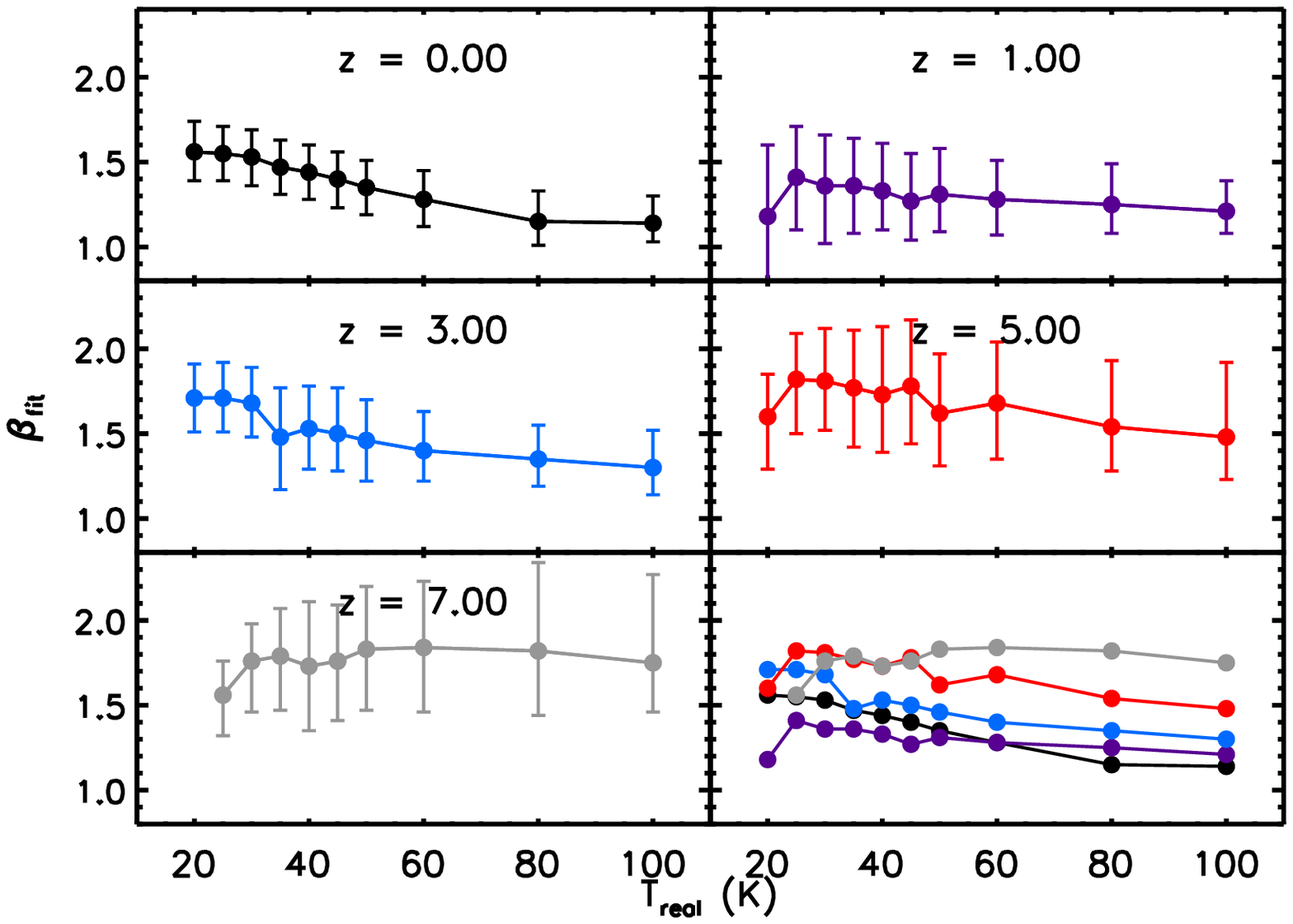}
\caption{
Same as Fig.~\ref{Fig_massfit_rawMAC}, but showing $\beta$ rather than \mfit.
}
\label{Fig_beta_rawMAC}
\end{center}
\end{minipage}
\end{figure*}

The synthetic photometry produced in the previous section has to be interpreted. We will obtain typical dust observables -- mass, temperature, emissivity index $\beta$ -- by fitting the synthetic fluxes with a typical model: a MBB with an power-law opacity plus a CMB correction for high-redshift objects, i.e. the product of Eq.~\ref{eq_mbb_simple} and Eq.~\ref{eq_CMBcorr_singleT}, where \tfit\ replaces T$_{\rm dust}$ \citep[see e.g.][]{daCunha+13}. As mentioned in Sect.~\ref{section_opacity_compare}, we use the $\kappa_0$ from \citet{James+02} -- $0.7\,{\rm cm}^2\,{\rm g}^{-1}$ at $\lambda_0 = 850\,\mu$m -- as representative of where opacity values from the literature tend to group (see also Fig.~\ref{Fig_Opacity_labvslit}).

\subsection{Fitting procedure}
\label{section_fit_procedure}

Our fit treats mass, temperature and $\beta$ as free parameters. In the remainder of the paper we use the terms \treal\ and \mreal\ for the temperature and dust mass used in the creation of the synthetic photometry, and the terms \tfit, \mfit\ and \bfit\ for the temperature, mass and $\beta$ obtained from the fits (note that there is no $\beta_{\rm real}$ because the opacity used for the synthetic SED is not a simple power law). It is assumed in our fits that the redshift of the source is known. 

The fit uses Markov Chain Monte-Carlo (MCMC) algorithms to carry out Bayesian inference. We fit the three parameters \tfit, \bfit\ and $\log_{10}$(\mfit) using flat priors with the following bounds: \tcmb(z) $<$ \tfit $<300\,$K, $0.5 <$ \bfit $< 4$ and $35 < \log_{10}($\mfit$) < 45$, where \mfit\ is in grams so that $\sim50\,{\rm M}_\odot < $ \mfit $< 5\cdot10^{11}\,{\rm M}_\odot$. Tests made using non-flat priors on \tfit\ and \bfit\ do not significantly improve the results, and in fact can significantly skew the temperature determination. 
We chose to run our fit with 30 walkers over 1000 steps, with a burn-in of 200. Test runs show that these numbers converge to the same results as longer runs using 100 walkers, 10$^4$ steps and a burn-in of 1000. The fiducial value and (asymmetric) uncertainties on each parameter are obtained from the \nth{16}, \nth{50} and \nth{84} percentiles of the MCMC final sampling, as in Sect.~\ref{section_mock_observations}. In this case the positive and negative error bars are not always similar, so we keep both the positive and negative uncertainties on the fit parameters.

To test the reliability of our fitting routine, we also performed fits on a ``test'' SED that has the same opacity as the fitting model: $\kappa = 0.7\,{\rm cm}^2\,{\rm g}^{-1}$ at $\lambda_0 = 850\,\mu$m, with a value of 1.5 for $\beta$. As shown in Appendix~\ref{Appendix_self-consistency}, test fits for single-temperature dust generally obtain correct values for \mfit, \tfit\ and \bfit. 

\subsection{Bias in recovered dust masses}
\label{section_fit_dustmass}

For the photometry using the raw opacity, fit results are shown in Fig.~\ref{Fig_massfit_rawMAC} (dust mass), \ref{Fig_Tfit_rawMAC} (temperature) and \ref{Fig_beta_rawMAC} (power law index $\beta$). The figures show a representative subset of the results as a function of both \treal\ and z. 
The most striking feature of these fits is that they overestimate the dust mass by a factor of $\sim 20$, independent of redshift. We would expect the ratio \mfit/\mreal\ to increase with \treal, since dust opacity increases with temperature; however, the trend is not readily visible in the data. While the median \mfit/\mreal\ increases from $\sim15$ at \treal$ = 20$~K to $\sim22$ at \treal$ = 100$~K; the error bars on \mfit\ -- especially at high temperature -- are comparable with, or larger than, this difference. We conclude that the temperature dependence of opacity has little impact on \mfit\ compared to the uncertainties of the fit itself. The uncertainty on the \mfit/\mreal\ ratio is dominated by the uncertainty on the \citet{James+02} $\kappa_0$ value itself, $0.7\pm0.2\,{\rm cm}^2{\rm g}^{-1}$. This gives a $\sim30$ per cent uncertainty on all determinations of \mfit/\mreal\ which, being systematic, affects the ratio at all temperature in the same way.

In contrast with the results for \mfit, \tfit\ remains consistent with \treal, although above $\sim 50\,$K the error bars on \tfit\ increase significantly. The sign of \tfit\ $-$ \treal\ depends on redshift, which is probably an indirect effect of z affecting the SED sampling: the same instrumental bands correspond to different (rest-frame) wavelengths at different redshifts.  
The values of \bfit\ all fall in a physically plausible interval of 1 -- 2. A tendency of \bfit\ to decrease with \treal\ can be observed for those redshifts where \tfit$>$\treal\ at high temperature; this is likely an effect of the degeneracy between T and $\beta$ in MBB fits. At high redshift, where $\beta$ is not well constrained, this tendency is no longer evident.

\begin{figure*}
\begin{minipage}{\textwidth}
\begin{center}
\includegraphics[width=.9\hsize]{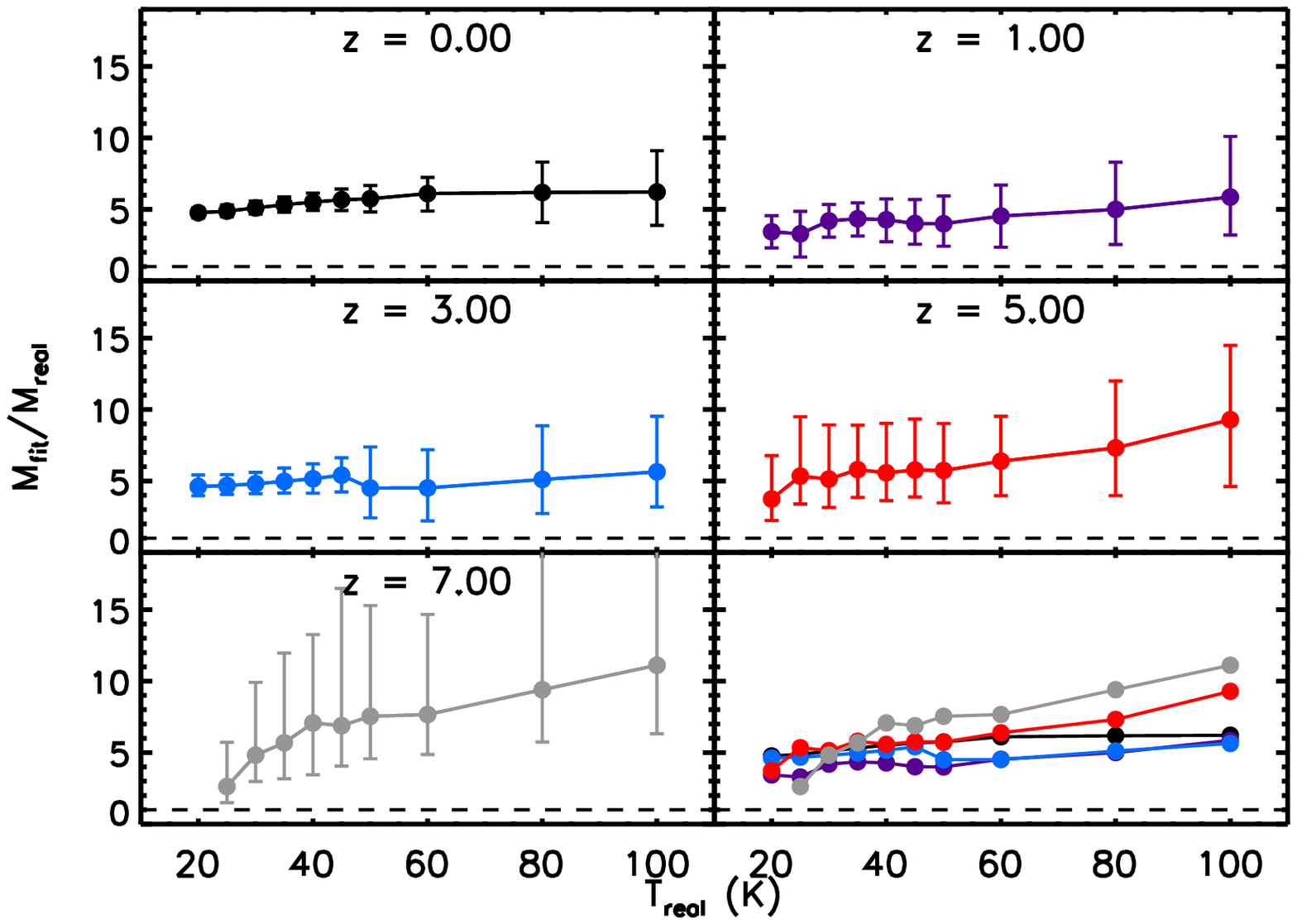}
\caption{
Same as Fig.~\ref{Fig_massfit_rawMAC} (\mfit\ plot) but including a correction on the opacity to account for grain fluffiness. 
}
\label{Fig_massfit_oc}
\end{center}
\end{minipage}
\end{figure*}

\begin{figure}
\begin{minipage}{0.5\textwidth}
\begin{center}
\includegraphics[width=.99\hsize]{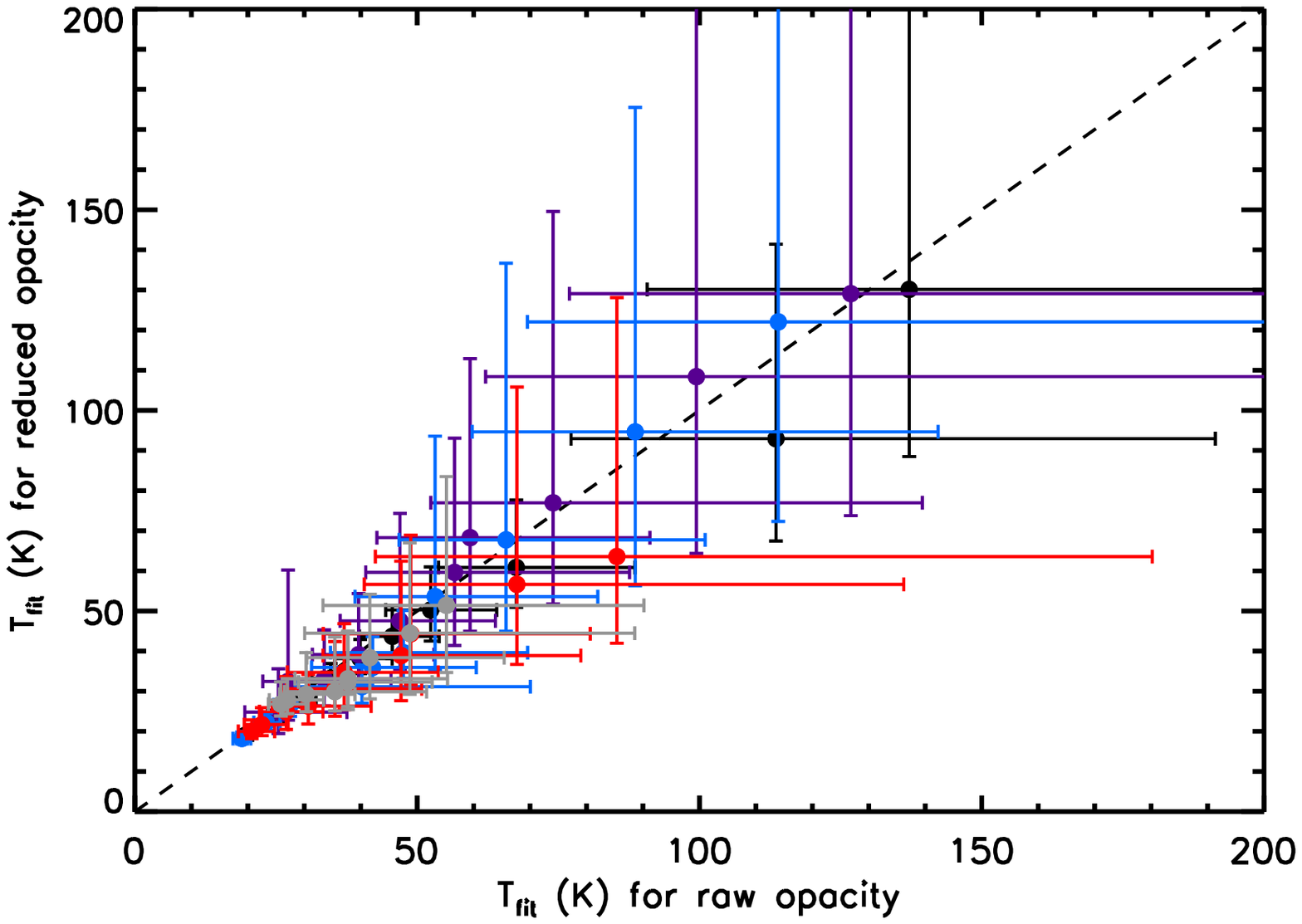}
\includegraphics[width=.99\hsize]{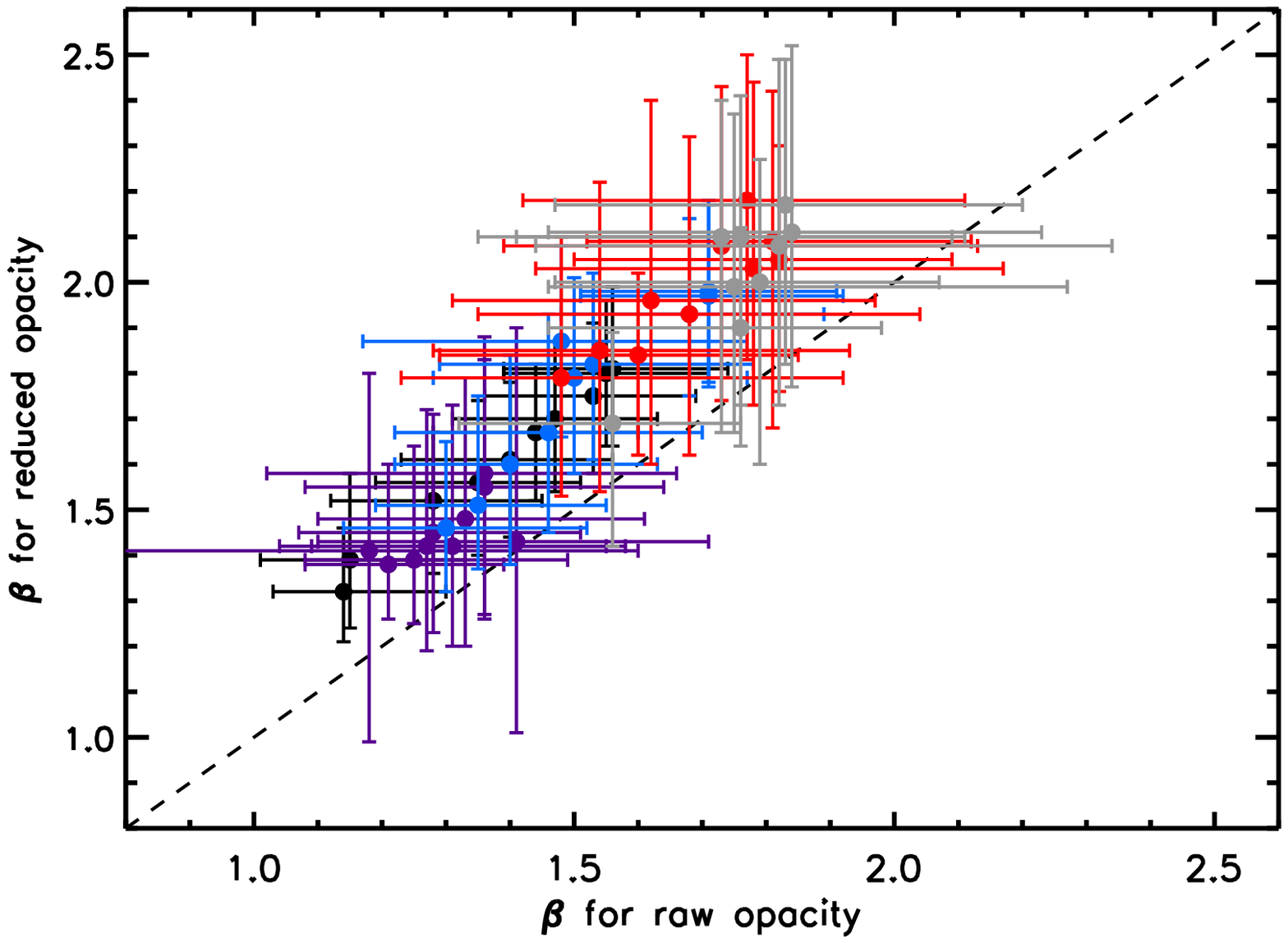}
\caption{
Applying the opacity correction does not influence \tfit. It increases \bfit\ slightly on average ($\sim 1\, \sigma$). The color code used is the same as Figs.~\ref{Fig_massfit_rawMAC} to \ref{Fig_massfit_oc}.
}
\label{Fig_TBcompare}
\end{center}
\end{minipage}
\end{figure}

The fit results for the reduced opacity photometry are shown in Fig.~\ref{Fig_massfit_oc} (\mfit) and \ref{Fig_TBcompare} (\tfit\ and \bfit, compared to the results for the raw $\kappa$). The value of \mfit, although smaller than in the case of raw opacity, remains 4 to 7 times higher than \mreal. Since these results are for the case with reduced opacity, they should be seen as a lower limit to \mfit, at least within our modelling framework. The opacity correction does not have a large influence on the other fit results (see Fig.~\ref{Fig_TBcompare}): the values for \tfit\ are indistinguishable from those Sect.~\ref{section_fit_dustmass}, and while the index \bfit\ is slightly higher in the reduced opacity case -- due to the fact that the larger correction was for carbon, which has a shallower dependence on $\lambda$ -- the difference is of the order of $1\sigma$. 

\subsection{Shorter-wavelength fit: two-temperature synthetic photometry}
\label{section_fit_multiT}

\begin{figure}
\begin{minipage}{0.5\textwidth}
\begin{center}
\includegraphics[width=.99\hsize]{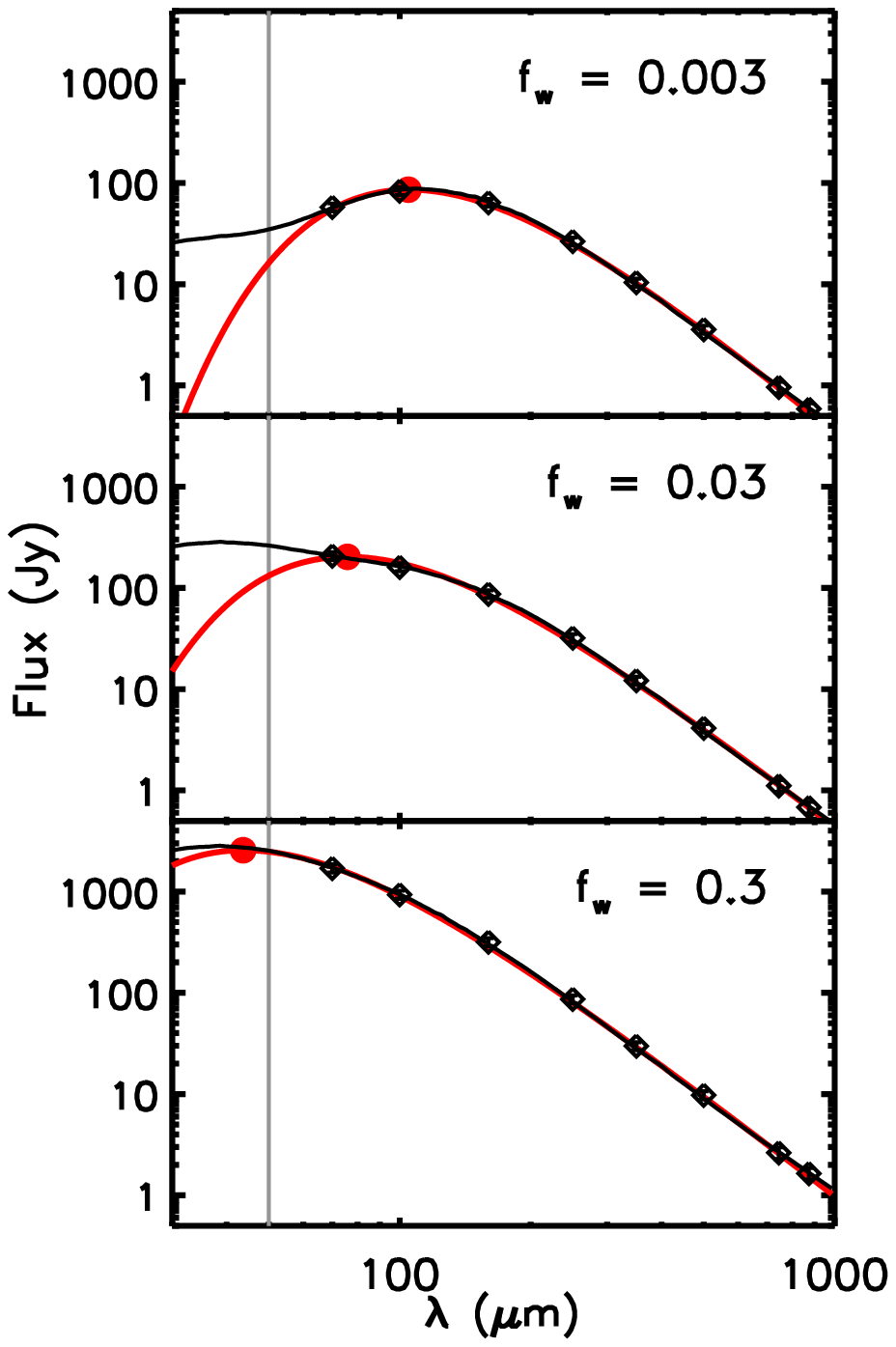}
\caption{
Dust emission SED for two-temperature dust, for increasing values of \fw\ (see text). The two-temperature SEDs are in black; the lozenges show the corresponding photometry and the red curves are the single-temperature fit to it. The position of the fit's peak is marked by a red disc. The grey vertical line shows the $\lambda = 50\,\mu$m limit. 
}
\label{Fig_SED_2T}
\end{center}
\end{minipage}
\end{figure}

As explained in Sect.~\ref{section_model_multiT}, if we want to fit shorter wavelengths ($\lambda \lesssim 100\,\mu$m, where the emission for warm dust is expected to peak), we need to add a warmer dust component. In this section we will examine the fit results on a mixture of $30\,$K and $100\,$K dust. Fig.~\ref{Fig_SED_2T} shows how the SED shape changes as the mass fraction of warm dust, \fw, increases.
The fit was performed in the same way as in Sect.~\ref{section_fit_dustmass}, but including bands with $\lambda \geq 50\, \mu$m \citep[the limit of validity for MBB fits as per][]{daCunha+13}. Due to the discrete nature of the photometric bands used, this does not necessarily mean that all fits include $\lambda \sim 50\,\mu$m (rest-frame) data. The SEDs to fit were calculated using the reduced opacity described in \ref{section_model_redmac}, and therefore our values for \mfit\ in this section represent a conservative estimate.

\begin{figure*}
\begin{minipage}{\textwidth}
\begin{center}
\includegraphics[width=.99\hsize]{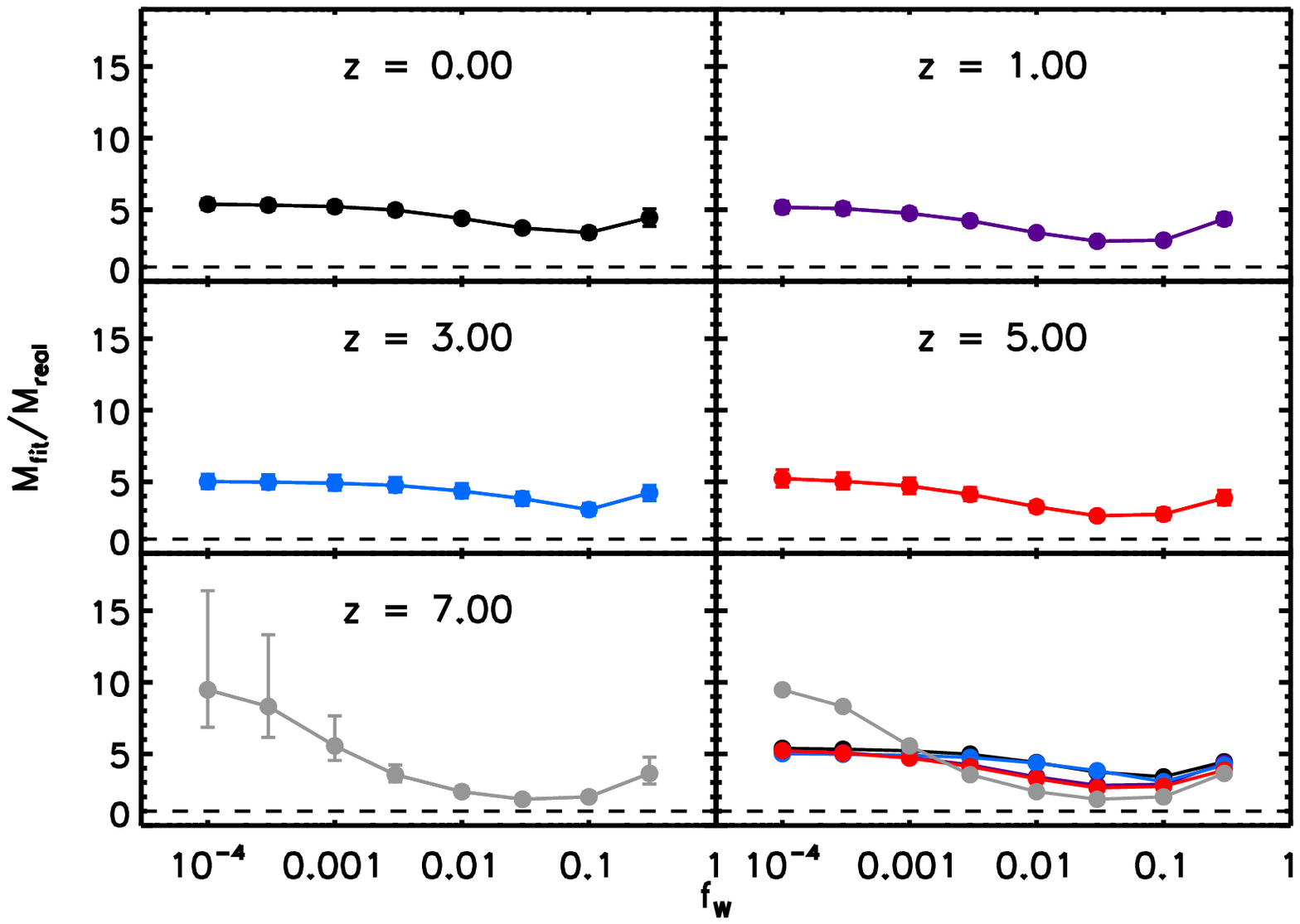}
\caption{
\mfit\ results of one-temperature fits to  the two-temperature SEDs, using bands with $\lambda \geq 50\,\mu$m. Showing results for the reduced-opacity dust. 
}
\label{Fig_Mfit_2T}
\end{center}
\end{minipage}
\end{figure*}

\begin{figure*}
\begin{minipage}{\textwidth}
\begin{center}
\includegraphics[width=.9\hsize]{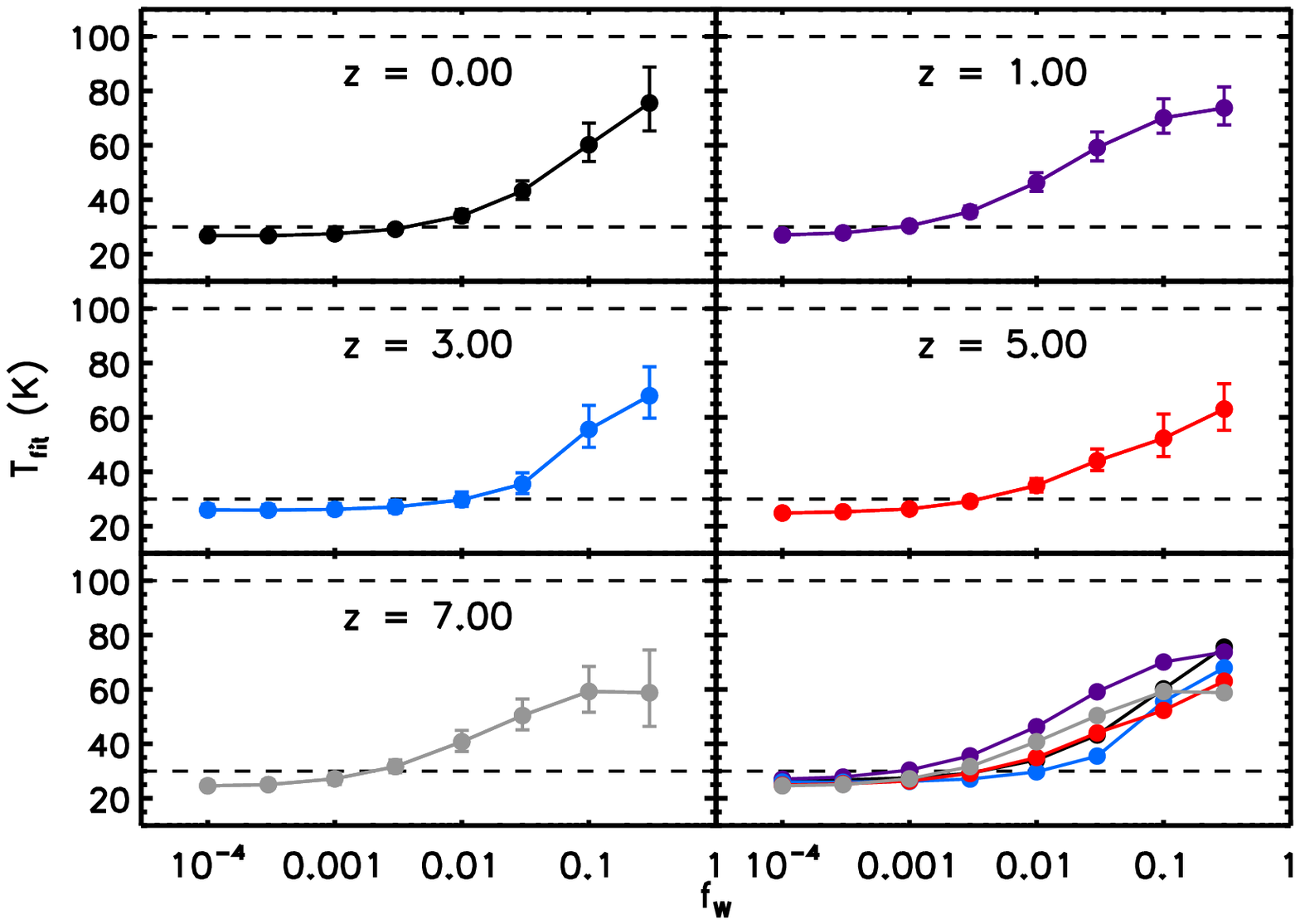}
\caption{
\tfit\ results of one-temperature fits to  the two-temperature SEDs, using bands with $\lambda \geq 50\,\mu$m. Showing results for the reduced-opacity dust. 
}
\label{Fig_Tfit_2T}
\end{center}
\end{minipage}
\end{figure*}

\begin{figure}
\begin{minipage}{0.5\textwidth}
\begin{center}
\includegraphics[width=0.99\hsize]{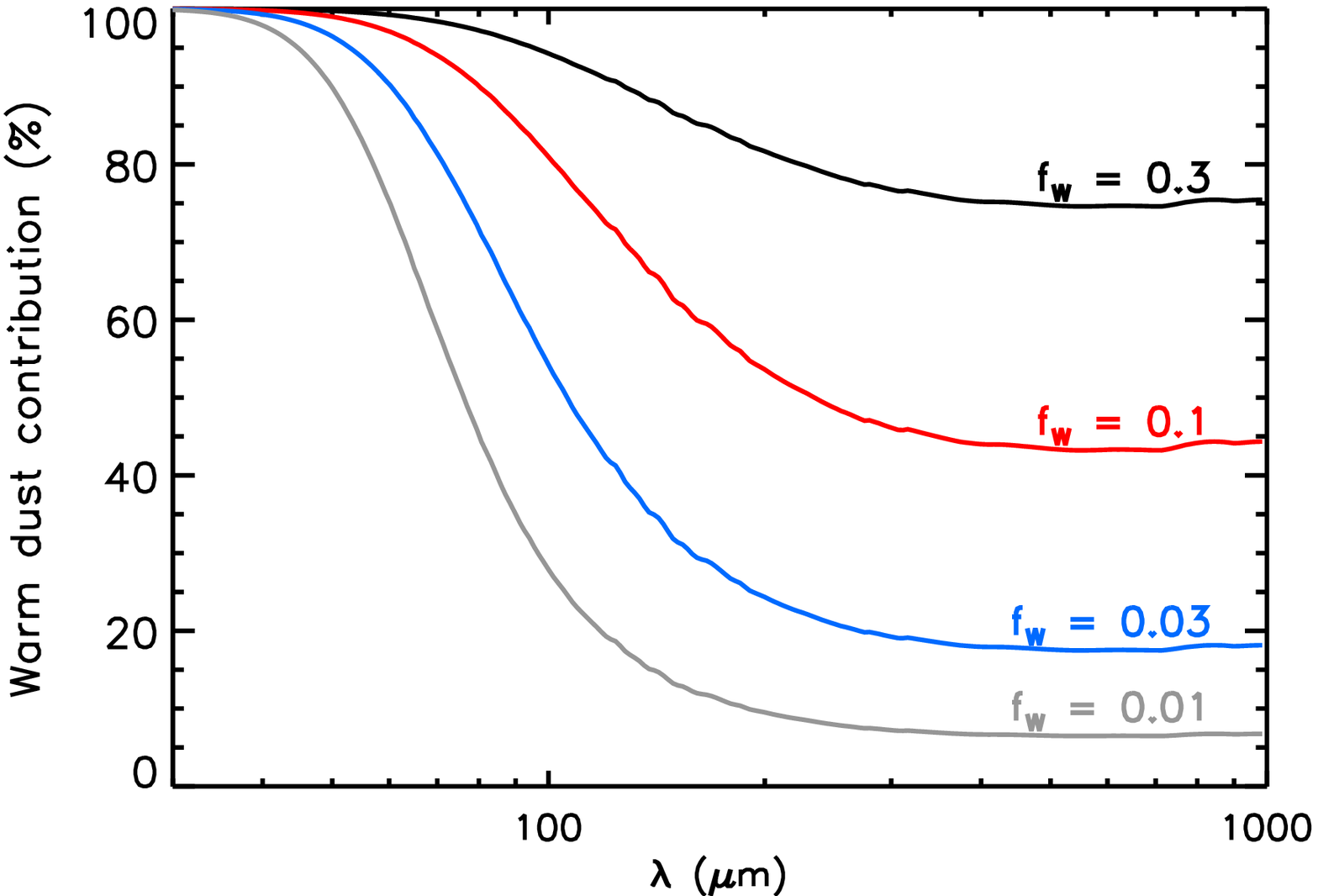}
\caption{
Contribution of the warm (100 K) dust component to the total SED as a function of wavelength and \fw. Showing results for the reduced-opacity dust. Note that the curves are not perfectly smooth, due to the small features present in the experimental opacities.
}
\label{Fig_warm_dust_contribution}
\end{center}
\end{minipage}
\end{figure}

The fit results are shown in Figs.~\ref{Fig_Mfit_2T} (\mfit) and \ref{Fig_Tfit_2T} (\tfit) as a function of \fw\ and z. Both \mfit\ and \tfit\ have much smaller error bars than in the case of single-temperature dust (Sect.~\ref{section_fit_dustmass}), likely due to the inclusion of shorter wavelengths in the fit. We can observe a general trend of \mfit\ first decreasing, then increasing again with \fw, which is an artefact of fitting the SED with a single-temperature model. For small values of \fw\ the results are very close to those for single-temperature 30 K dust, but as \fw\ increases, the SED increases at short wavelength first (Fig.~\ref{Fig_warm_dust_contribution}), which results in a higher \tfit\ and -- to preserve the long-wavelength photometry -- a lower \mfit. After \fw\ increases above a few percent, the long-wavelength part of the SED increases as well, which results in a higher \mfit. Note that the lowest \mfit\ is still about twice the value of \mreal: even in the most favorable scenario where we use reduced opacity and a warm dust component lowering \mfit, the fit overestimate dust masses.

While in most cases \mfit/\mreal\ tends to $\sim$5 for small \fw, i.e. the same value as for single-temperature 30 K dust, the case of z = 7 is peculiar and shows a higher value of $\sim$10. This is because our fit slightly underestimates \tfit\ (Fig.~\ref{Fig_Tfit_2T})\footnote{The difference between \tfit\ and \treal\ is likely a consequence of fitting a power-law opacity to dust with a non-power-law $\kappa$. If so, the sign and magnitude of \tfit\ - \treal\ depend on the shape of $\kappa(\lambda)$ and therefore, ultimately, on dust composition.} and \tfit$\sim$\tcmb\ (which for z = 7 is 21.8 K), so the correction for the flux lost to CMB subtraction (see Sect.~\ref{section_model_CMB}) is overestimated. Since cases where \tfit\ $\sim$ \tcmb\ are unlikely at high redshift, we do not expect this effect to have practical consequences on dust mass fits.

\subsection{Simulating very high-z fits: two-band photometry}
\label{section_fit_2bands}

The observation and mass determination of dust in very high-redshift galaxies (z$\,\gtrsim 6 - 7$) is an important endeavor in current astrophysics: dust emission is used as a tracer of star formation and gas mass, and therefore it is an important messenger of galaxy evolution in the Universe. Another source of interest for high-z dust is the so-called ``dust budget crisis'': interstellar dust is thought to form mainly in the atmosphere of evolved stars but, since at z~=~7 the Universe is only $\sim700$~Myr old, the dust masses derived at high redshift are too large to be explained that way unless one invokes unrealistic star formation rates \citep[e.g.][]{Morgan&Edmunds03, Rowlands+14, Watson+15}. In an attempt to resolve this tension, supernova ejecta \citep[e.g.][]{Dwek+14} and dust growth in the interstellar medium \citep{Micha+15, Mancini+15, Popping+17} have been suggested as additional dust sources, but there are doubts on the efficiency of both these pathways \citep{Ginolfi+18, Ferrara+16}. The dust budget crisis applies not only to the early Universe, but to the local Universe as well \citep[e.g.][]{Temim+15, Srinivasan+16}.

Unfortunately, the data for very high redshift galaxies tend to be very sparse, with sometimes as few as one or two photometric bands per galaxy. This is the case for two the most prominent examples of dusty galaxies at z $> 7$: A1689-zD1 \citep[z = 7.5,][]{Watson+15,Knudsen+17} and B14-65666 \citep[``Big Three Dragons''; z = 7.15,][]{Hashimoto+19}. Since there are too few data points for a regular MBB fit, the dust mass must be recovered with an alternative method, described in \citet{Knudsen+17} and \citet{Hashimoto+19}: a value of $\beta$ is chosen, which is then used to calculate a theoretical curve for the flux ratio of the two bands as a function of temperature, under the assumption that the emission follows a MBB and taking the CMB background into account (Sect.~\ref{section_model_CMB}). The intersection of this curve with the observed flux ratio gives the estimate of \tfit. Finally, given an assumed dust opacity, the value of \mfit\ is then the one that correctly reproduces the galaxy's IR luminosity, together with the $\beta$ and \tfit\ determined above.

Given the importance of dust mass determination at high redshift, we decided to test this mass fitting technique as we did for the ``regular'' MBB fit. We fit two bands of our synthetic dust photometry employing the same technique described above, with one modification -- since we are not interested in the IR luminosity as a quantity, after determining \tfit\ we obtain \mfit\ from a minimum-$\chi^2$ fit to the photometry, where the dust mass is the only free parameter. We decided to focus on the most used ALMA bands -- band 7, 6 and 3 -- meaning that we tested three two-band combinations: bands 7 and 6, bands 6 and 3, and bands 7 and 3. As can be seen in Table \ref{tab:restwls}, the central rest-frame wavelength for these bands is always longer than $100\,\mu$m, so we are justified in using a single-temperature model. Note that neither \citet{Knudsen+17} nor \citet{Hashimoto+19} used band 3: we use this band not to test preexisting results, but to verify its viability for these kinds of fits. We use the same set of $\beta$ values as \citet{Knudsen+17} and \citet{Hashimoto+19}, i.e. 1.5, 1.75 and 2. 

\begin{figure*}
\begin{minipage}{\textwidth}
\begin{center}
\includegraphics[width=.9\hsize]{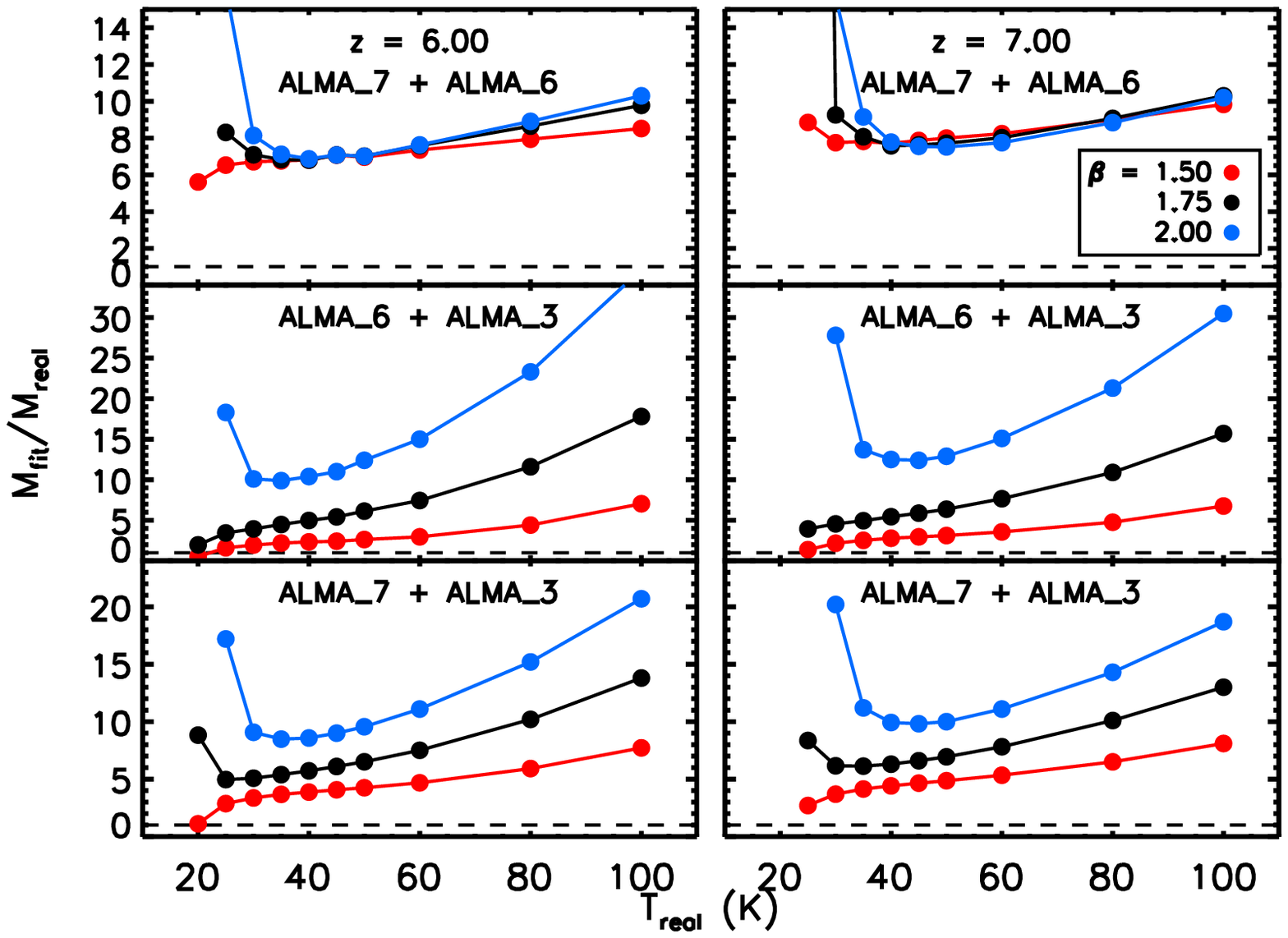}
\caption{
\mfit\ results for the ``two-band fit'' described in the text, for z = 6 (left) and 7 (right). The horizontal dashed line corresponds to \mfit/\mreal$\,= 1$. Note that the vertical scale is different in each row.
}
\label{Fig_2band_Mfit}
\end{center}
\end{minipage}
\end{figure*}

\begin{figure*}
\begin{minipage}{\textwidth}
\begin{center}
\includegraphics[width=.9\hsize]{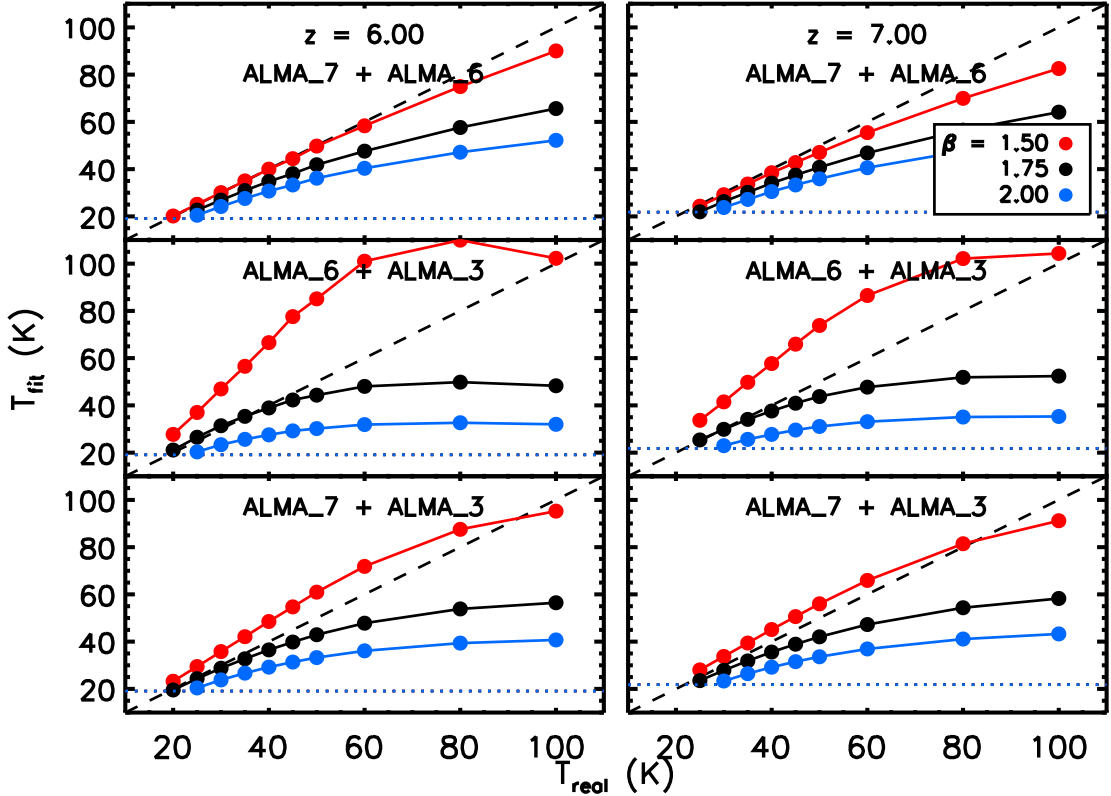}
\caption{
\tfit\ results for the ``two-band fit'' described in the text. The horizontal dotted line corresponds to \tcmb\ at that redshift.
}
\label{Fig_2band_Tfit}
\end{center}
\end{minipage}
\end{figure*}

\begin{table}
\begin{minipage}{0.5\textwidth}
\caption{Rest wavelengths for the filters used in the two-band fits.}
\centering
\begin{tabular}{rcccc}
    Band: & ALMA 7 & ALMA 6 & ALMA 3 \\
    \hline
    Central $\lambda\,(\mu$m), z = 0 & 872.76 & 1286.66 & 3074.79 \\
    z = 6 & 124.68 & 183.81 & 439.26 \\
    z = 7 & 109.09 & 160.83 & 384.35 \\
    \hline
\end{tabular}
\label{tab:restwls}
\end{minipage}
\end{table}

Fig.~\ref{Fig_2band_Mfit} and \ref{Fig_2band_Tfit} show \mfit\ and \tfit, respectively, for reduced-opacity dust at  z = 6 and 7. The main result from the previous sections -- that masses are usually overestimated -- is still true in most cases; the choice of bands used, however, is very influential. For ALMA bands 7 and 6, \mfit/\mreal\ is independent of $\beta$ and has a value of $\sim 6 - 10$ (slightly higher than for the reduced-opacity standard fit), with a shallow dependence on \treal. When fits include band 3, however, \mfit/\mreal\ is larger for high $\beta$, and can increase by over one order of magnitude as $\beta$ goes from 1.5 to 2. Since band 3 has the longest wavelength, our tentative conclusion is that the choice of $\beta$ in two-band fits becomes more important when one samples the long-wavelength end of the SED. This result is reminiscent of that of \citet{Bianchi_13}, who examined the combined effects of $\beta$ and $\kappa_0$ in MBB fit, and found that using too high a $\beta$ leads to overestimating dust masses, unless $\kappa_0$ is decreased accordingly.

At low temperature (\treal\ $\lesssim 30$ K) the fit results behave peculiarly and deserve a closer look. For $\beta = 2$, \mfit/\mreal\ is much higher than one could expect from the rest of the plot; for $\beta = 1.5$, on the other hand, \mfit/\mreal\ is especially low and can be close to unity. 
Both cases can be explained by a biased value of the CMB correcting factor $C_{\rm CMB}$ (Eq.~\ref{eq_CMBcorr}), which is very sensitive to small variations of \tfit\
when \tfit\ $\sim$ \tcmb. As one can see in Fig.~\ref{Fig_2band_Tfit}, fits with $\beta = 2$ have indeed the lowest \tfit, always lower than \treal\ and sometimes close to the CMB temperature. Fits with $\beta = 1.5$, on the other hand, can have a \tfit\ significantly higher than \treal, especially for the ALMA 6 + ALMA 3 combination. Nonetheless, since typical dust temperatures at z~$>$~6 are higher than 30 K \citep[e.g.][]{Watson+15, Hashimoto+19}, these low-temperature effects are unlikely to play a role in actual high-z fits.

In conclusion, our attempts at reproducing the two-band, high-z fits by \citet{Knudsen+17} and \citet{Hashimoto+19} using ALMA bands 6 and 7 show the same bias as the mass fits from previous sections or higher (\mfit/\mreal$\sim 6 - 10$ as opposed to the $\sim 4 - 7$ from Sect.~\ref{section_fit_dustmass}). We have also shown that a two-band fit including longer wavelengths, such as band 3, would be much less efficient at constraining dust masses.

\subsection{Alternative opacities from the scientific literature}
\label{section_fit_alternatives}

Although the opacity from \citet{James+02} is a good representative of ``classic'' FIR/submm opacity, as explained in Sect.~\ref{section_opacity_compare}, one may be interested in seeing what results would be obtained with other, more commonly used models. In this section we offer a quick comparison with two other model: \citet{Draine&Li07} (hereafter \citetalias{Draine&Li07}) and \citet{Compiegne+11} (hereafter \citetalias{Compiegne+11}).

As explained in \citet{Bianchi_13}, the \citetalias{Draine&Li07} opacity can be approximated as a power law with $\beta = 2.08$ and $\kappa_0 = 4\, {\rm cm}^2\, {\rm g}^{-1}$ at $\lambda_0 = 250\, \mu$m. This corresponds to a $\kappa$ of $0.3\, {\rm cm}^2\, {\rm g}^{-1}$ at $850\, \mu$m. A free-$\beta$ fit using this opacity would result in a higher \mfit/\mreal\ than obtained in the previous sections by a factor of 7/3. 
This is not entirely surprising, since the \citetalias{Draine&Li07} model has been noted to overestimate dust masses because of its very low FIR opacity \citep{Planck_XXIX_DL07opacity, Fanciullo+15}. One additional issue is that the approximate \citetalias{Draine&Li07} $\beta$ is higher than the best-fit $\beta$ for our synthetic SEDs (Fig.~\ref{Fig_beta_rawMAC}). As pointed out in Sect.~\ref{section_fit_2bands}, a wrong value of $\beta$ can influence \mfit/\mreal\ just as a wrong value of $\kappa_0$; e.g. using too high a $\beta$ results in overestimating \mfit/\mreal. This means that the aforementioned increase by factor 7/3, without considering $\beta$ effects, is in fact a conservative estimate.

In the case of \citetalias{Compiegne+11}, the power-law approximation has $\beta = 1.91$ and $\kappa_0 = 5.1\, {\rm cm}^2\, {\rm g}^{-1}$ at $\lambda_0 = 250\, \mu$m, resulting in a $850\, \mu$m opacity of $0.5\, {\rm cm}^2\, {\rm g}^{-1}$. The difference in $\kappa_0$ alone means that the \mfit/\mreal\ from sections \ref{section_fit_dustmass} to \ref{section_fit_2bands} would be 40\% higher if we had used the \citetalias{Compiegne+11} opacity. This model's $\beta$ of 1.91 is also higher than the typical \bfit, meaning that this, too, is a conservative estimate.
In conclusion, using either the \citetalias{Draine&Li07} or \citetalias{Compiegne+11} dust models would result result in a greater \mfit/\mreal\ value than using \citet{James+02}, though more so for \citetalias{Draine&Li07}.

The comparison with \citetalias{Compiegne+11} is also interesting for the reason that this model has one material in common with our synthetic SEDs: BE carbon, although it uses the single-temperature (n, k) data from \citet{Zubko+96} rather than the temperature-dependent MAC from \citetalias{Mennella+98}.
\footnote{
Although the \citet{Colangeli+95} materials for which \citet{Zubko+96} calculate (n, k) are supposedly the same studied by \citetalias{Mennella+98} at cryogenic temperatures, the long-wavelength MAC in \citet{Colangeli+95} does not match the \citetalias{Mennella+98} MAC for the same material at any temperature, except as a first-order approximation. Therefore, the comparison between \citetalias{Mennella+98} and the  \citet{Zubko+96} refractive index is not straightforward.
} 
One might wonder, therefore, why the \citetalias{Compiegne+11} $\kappa$ is so much lower than even the reduced opacity we use. There are two main reasons for this:
\begin{itemize}
    \item We use a different method to convert BE opacity from the raw experimental value to the ``reduced'' version. Our method was chosen partly for computational simplicity, but it also underlies a different set of assumptions about grains properties (see below).
    \item \citetalias{Compiegne+11} use the same silicate as \citet{Li&Draine01}. The FIR/submm opacity of these silicates is extrapolated from shorter wavelengths and it has been shown to underestimate experimentally-measured silicate $\kappa$ (see e.g. \citetalias{Demyk+17B}).
\end{itemize}
To expand on the first reason mentioned, the effects of aggregate formation on opacity are strongly dependent on the method used for the calculation \citep[e.g.][]{Krueg&Sieb_94, Stognienko+95, Ysard+18}.
The opacity in \citetalias{Compiegne+11} is obtained via Mie theory from the BE (n, k) data by \citet{Zubko+96}, derived from the experimental MAC using a modified Continuous Distribution of Ellipsoids \citep[mCDE, adapted from the CDE in][with increased weighting for elongated particles]{Boh+Huf83}. The \citetalias{Compiegne+11} model assumes that grains are compact and spherical. Compactness is a common assumption for the high-latitude diffuse ISM, but not necessarily true over an entire galaxy, as discussed in Sect.~\ref{section_model_redmac}. Furthermore, it is known that interstellar grains are not spherical even in the diffuse ISM, and one effect of non-sphericity is to enhance FIR/submm opacity \citep[e.g.][]{Hild&Drago_95,Guillet+18}. Therefore, one may reasonably expect the \citetalias{Compiegne+11} model to underestimate the actual FIR/submm opacity of interstellar dust. The question of whether our own method for ``reducing'' $\kappa$ (see Sect.~\ref{section_model_redmac}) may instead overestimate opacity is legitimate, but to test it would require to recover (n, k) data from the cryogenic \citetalias{Mennella+98} MACs, and is therefore outside of the scope of this paper. We underline again that a separate-grain dust model is not necessarily a more realistic choice than an aggregate model when averaging over a galaxy, although most works that derive dust masses in galaxies use diffuse-ISM, spherical-grain dust models with FIR opacities in the same order of magnitude as \citetalias{Compiegne+11} \citep[e.g.][]{Berta+16, Nersesian+19, Aniano+20, DeLooze+20}. If a significant fraction of dust emission comes from aggregates, the results in this kind of work would be biased.

\section{Astrophysical implications}
\label{section_astro_implications}

Our main conclusion is that current dust mass fits could be overestimating dust masses by a factor of $\sim2$ to $\sim 20$, depending on the physical properties and temperature distribution of interstellar grains; e.g., the presence of a warm dust component or the use of reduced opacity -- meant to represent single-grain dust, rather than the aggregates found in laboratory conditions -- result in lower dust masses.
The measurement of interstellar dust masses is essential to many astrophysical endeavors. For instance, dust mass is used as a gas mass tracer via the dust-to-gas ratio, which is assumed to scale with the metallicity, although this relationship breaks down at low metallicity \citep{Remy-Ruyer+14}. Then the gas density/mass is assumed to scale with the star formation rate via the Kennicutt-Schmidt law, so that dust mass measurements at a range of redshifts probe the star formation history of the Universe. Therefore, our results have far-reaching implications.

An example of these implications is the so-called ``dust budget crisis'' introduced in Sect.~\ref{section_fit_2bands}: the dust masses currently estimated at z $>$ 5 are not compatible with standard dust production channels, and require an overhaul in our models of the initial mass function for star formation, of supernova production rates, or of dust growth in the ISM. Overall, the dust production rate would need to increase by one to two orders of magnitudes, as shown by \citet{Rowlands+14}.
The growth of dust grains through accretion in the interstellar medium has been proposed as a solution \citep[e.g.][]{Micha+15, Mancini+15, Popping+17}, but there are doubts on the efficiency of accretion at high z, where high dust temperatures due to the CMB (see Sect.~\ref{section_model_CMB}) keep the desorption timescale for accreted materials short \citep{Ferrara+16}.
The dust budget crisis is not only a problem at high redshift; it is observed, e.g., in the Magellanic Clouds (SMC, LMC). As explained in \citet{Srinivasan+16} using the dust mass fits by \citet{Gordon+14}, the dust replenishment timescale in the SMC from stellar sources alone is expected to be larger than the dust destruction timescale and, in the worst-case scenario, longer than the lifetime of the Universe. Similarly, the ratio between the best LMC dust mass estimate by \citet{Gordon+14} and the dust injection estimates by \citet{Riebel+12} result in a LMC replenishment timescale of $34 \pm 8$ Gyr, exceeding the age of the Universe. Both the high-redshift and the local Universe, therefore, show a dust budget crisis that could be alleviated -- and, in the best case scenario, fully resolved -- if the actual dust masses turned out to be lower than currently estimated, as our results suggest. More specifically, \citet{Rowlands+14} mention that dust opacity needs to be increased by just a factor of 7 to solve the high-redshift crisis (provided dust destruction by SNe is not efficient); in the LMC, the aforementioned replenishment timescale would decrease to less than 2 Gyr if the dust mass were decreased by a factor of 20.

One caveat on our findings is that the experimental opacities are influenced by the size, shape and structures of the grains studied (see Sect.~\ref{section_model_redmac}). While we can neglect grain size, being in the Rayleigh regime, the fluffy aggregates typical of lab materials have a higher opacity per unit mass than isolate grains. Our reduced opacity attempts to correct for aggregates, but the effect of clustering on opacity are very model-dependent (see also Sect.~\ref{section_fit_alternatives}). However, it is debatable to what point opacity reduction is needed when modelling the SED of a full galaxy, which is bound to contain aggregates.

Another issue to consider is that, due to the lack of constraints on the interstellar dust composition, the materials we employ may not be the same that compose actual interstellar dust, so they may have a higher opacity than that of the actual components. However, our findings are not limited to a composition of 70 per cent reduced Mg$_{0.7}$Fe$_{0.3}$SiO$_3$ silicate (E30R) and 30 per cent amorphous carbon (BE); fit results are qualitatively the same for every combination of materials we tried, as mentioned in Sect.~\ref{section_model}. If we assumed that a difference in opacity between interstellar and laboratory materials contributes to the high values of \mfit/\mreal\, this would require that interstellar dust have a lower opacity than any material of similar stoichiometry we have been able to produce so far.
This said, it is important to look for independent confirmation for our low dust masses, by checking for instance if our revised masses are consistent with other dust tracers, such as dust extinction and elemental depletion. 

The comparison of dust mass estimates from emission and extinction in the local Universe gives ambiguous results. Traditional models are capable of fitting dust extinction and emission simultaneously in the Milky Way \citep{Fanciullo+15}, but on the other hand their FIR/submm opacity is too low to justify the polarization fraction in emission observed by {\it Planck} \citep{Guillet+18}. The comparison of FIR emission and NIR extinction in M31 also suggests that the FIR opacity of most dust models is too low, at least compared to the NIR/optical opacity \citep[][and refs. therein]{Whitworth+19}.

Dust masses estimated from metallicity and elemental depletions are generally considered to be consistent with the values (and therefore the opacities) of preexisting dust models. For instance, the \citet{James+02} opacity we used in the present work was found by calibrating two-temperature MBB fits of galactic SEDs with observed elemental depletion, and more recent reanalyses prefer, if anything, even lower opacities \citep[e.g.][]{Clark+16}. However, large uncertainties remain in this kind of mass determination. For instance, \citet{Kew&Ell_08} find that the choice of calibration can introduce a scatter of 0.7 dex (or about a factor of 5) in the determination of the absolute metallicity of outer galaxies.

Considering the ambiguous results obtained when comparing our dust masses to those obtained with other dust tracers, further work is needed to reconcile the different estimates. In particular, it will be necessary to review the assumptions used in the cross-comparison of dust tracers, to see whether other systematic errors may be present.

\section{Conclusions and future work}
\label{section_conclusion}

We built a model of FIR/submm dust emission using the opacity of dust analogues measured in the lab \citepalias{Mennella+98, Demyk+17A, Demyk+17B}. We then fit the synthetic SEDs with a model typically used by observers, a MBB with a power-law opacity, to test whether the fit could recover the mass and properties of the simulated dust. We used the opacity from \citet{James+02} -- 0.7$\pm$0.2 cm$^2\,$g$^{-1}$ at 850 $\mu$m -- as representative of typical dust models. We have found that these fits overestimate dust masses by a factor of $10 - 20$ or $2 - 5$, depending on the assumptions on grain structure (porous or compact, respectively). This is comparable to the excess dust mass observed in the ``dust budget crisis'' in both the local and the high-redshift Universe. These large fit masses are mainly due to the higher opacity of lab materials compared to that assumed in typical MBB models. The other large difference between experimental and extrapolated dust opacity -- the increase of $\kappa$ with temperature --  has a comparatively minor effect in the range of temperature we studied ($\leq 100$ K). We were unable to estimate the temperature-dependent increase of opacity since it appears to be smaller than the uncertainties on the MBB fits themselves, especially at high temperature. In contrast with the mass fit results, the fitted temperatures are consistent with \treal, and \bfit\ remains within a realistic range of values. 

It should be noted that, while \mfit/\mreal\ is model-dependent, our main result -- that \mfit\ is significantly higher than \mreal\ -- remains valid for all the combinations of redshift, temperature distributions, fitting methods and material compositions that we tried. Therefore, while the composition of interstellar dust may differ from the ones we chose, that would not solve the central issue of this work. No matter which explanation of the high \mfit/\mreal\ one takes to be closer to correct, our results show that two branches of dust astrophysics -- model fitting and laboratory analyses -- have worked independently of each other to give contradictory results, and this contradiction needs to be resolved if we want to understand interstellar dust.

Comparison with other dust tracers, such as dust extinction and elemental depletion, gives ambiguous results, as some are consistent with significantly reduced dust masses while others are apparently not. This underlines the need for a more complete cross-checking of dust mass estimates to search for possible systematics. Furthermore, the model used to produce synthetic SEDs in this paper is quite simple, and follow-up works will endeavor to take steps towards greater realism, such as relaxing the assumption of optically thin sources, using more complete temperature distributions, and calculating the dust mass absorption coefficient from the refractive index $(n,\,k)$ of the materials to have a better control over grain shape effects.

\section*{Data availability}

The code used in this article, together with the instructions on how to reproduce our results, is available in Github at \url{https://github.com/lfanciullo/Fanciullo_etal_2020_dust_mass_systematics}.

The data on silicate opacity are available in the Solid Spectroscopy database infrastructure (SSHADE), at
\url{https://www.sshade.eu/doi/10.26302/SSHADE/STOPCODA}.

The data on carbon opacity were provided by Vito Mennella (vito.mennella [at] inaf.it) by permission. Data will be shared on request to the corresponding author with permission of Vito Mennella.

\section*{acknowledgments}
LF wants to thank Karine Demyk and Vito Mennella for making available the data sets underlying the plots in their publications, James Simpson for his valuable advice on high-redshift photometry and for helping with modelling, and Harald Mutschke and Jonas Greif for useful discussions on the properties of dust analogues. Jonathan Marshall, as the reproducibility officer of the ASIAA interstellar and circumstellar matter group, provided invaluable help in testing the online scripts provided with the article.
Jennifer Karr has helped overcoming obstacles in the coding and Alfonso Trejo provided helpful insight on how ALMA bands work.

This research has been financially supported by the Ministry of Science and Technology of Taiwan under grant numbers MOST104-2628-M-001-004-MY3 and MOST107-2119-M-001-031-MY3, and by Academia Sinica under grant number AS-IA-106-M03.

SS acknowledges support from UNAM-PAPIITProgramme IA104820.

This research has made use of the SVO Filter Profile Service \url{http://svo2.cab.inta-csic.es/theory/fps/} supported from the Spanish MINECO through grant AYA2017-84089 \citep{SVOsite1, SVOsite2}.

This research has made use of IDL code downloaded from Chris Beaumont's IDL library \citep{site_medabsdev} and from the Heliodocs website \citep{site_percentiles}.

Our fit was implemented using the PYTHON packages Astropy \citep{Astropy_paper} and  {\tt EMCEE} \citep{emcee_paper}.




\bibliographystyle{mnras}
\bibliography{biblio_opacity}



\appendix

\section{Fit self-consistency test}
\label{Appendix_self-consistency}

\begin{figure*}
\begin{minipage}{\textwidth}
\begin{center}
\includegraphics[width=.9\hsize]{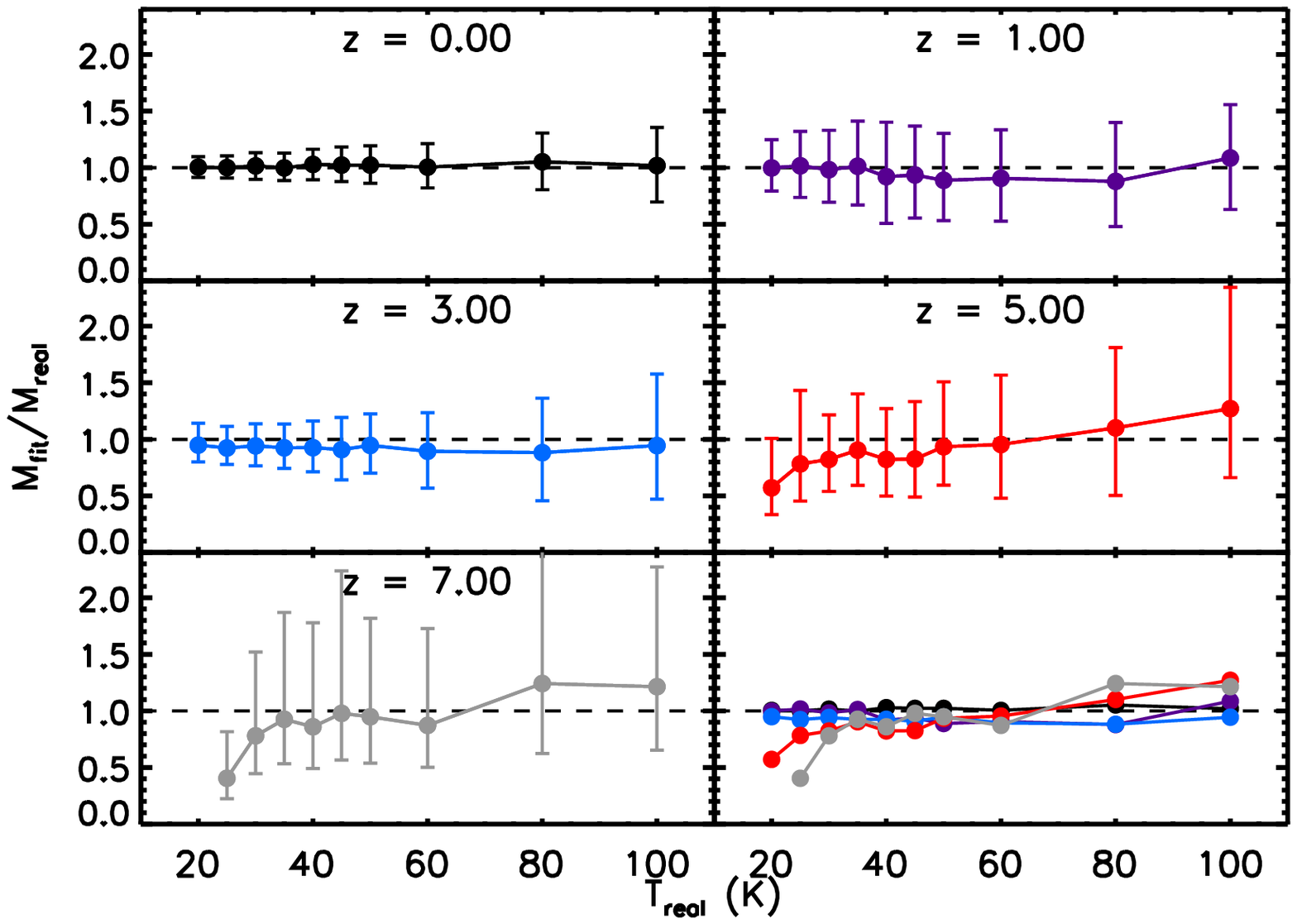}
\caption{
Mass fit results for our test model (MBB with $\kappa_0 = 0.7\,{\rm cm}^2\,{\rm g}^{-1}$ at $\lambda_0 = 850\,\mu$m, $\beta = 1.5$) as a function $T_{\rm real}$ and $z$ (compare Fig.~\ref{Fig_massfit_rawMAC}. The equality line \mfit/\mreal\ = 1 is shown as a dashed line. 
}
\label{Fig_massfit_MBBtest}
\end{center}
\end{minipage}
\end{figure*}

\begin{figure*}
\begin{minipage}{\textwidth}
\begin{center}
\includegraphics[width=.9\hsize]{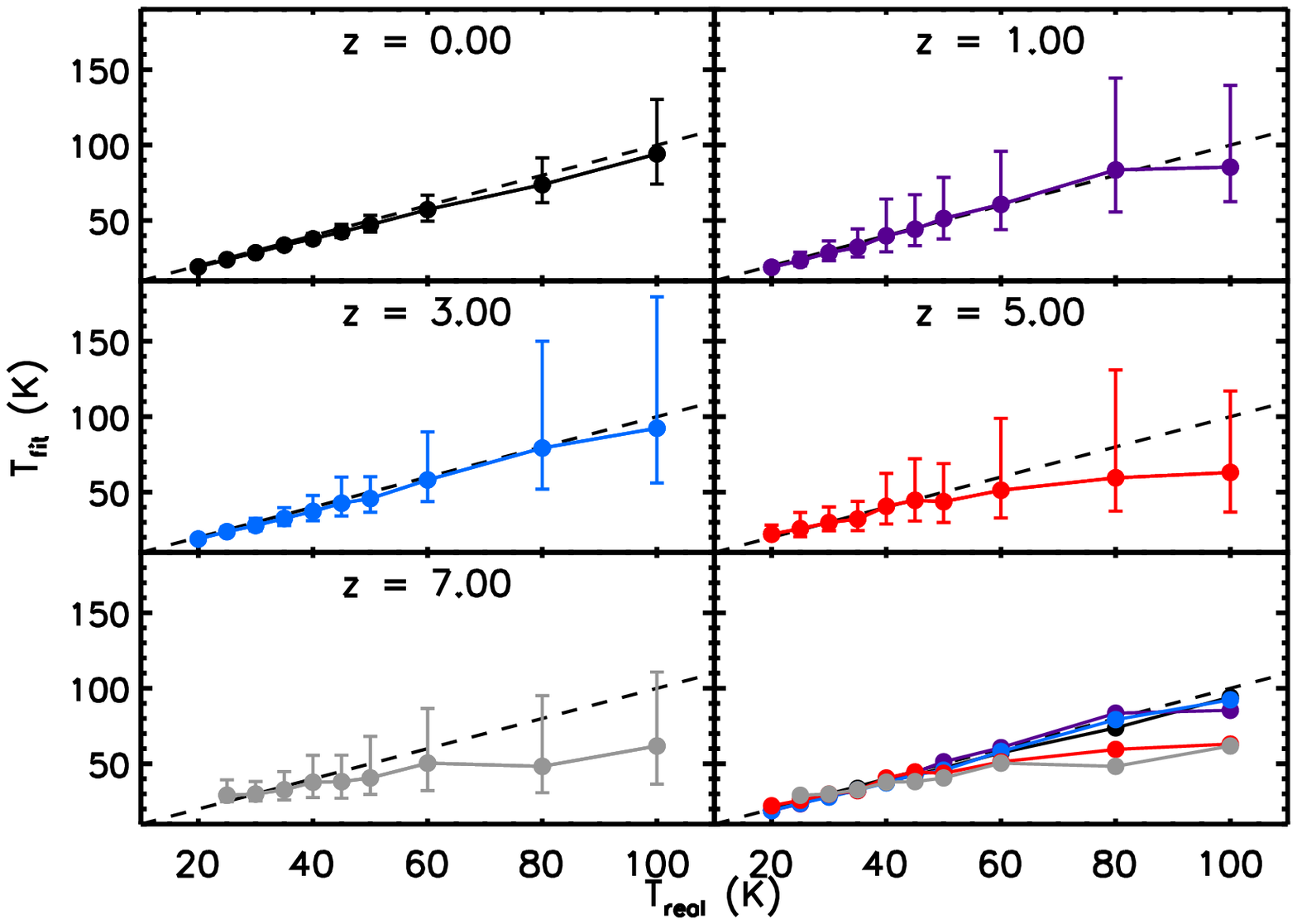}
\caption{
Same as Fig.~\ref{Fig_massfit_MBBtest}, but showing \tfit\ rather than \mfit.
}
\label{Fig_Tfit_MBBtest}
\end{center}
\end{minipage}
\end{figure*}

\begin{figure*}
\begin{minipage}{\textwidth}
\begin{center}
\includegraphics[width=.9\hsize]{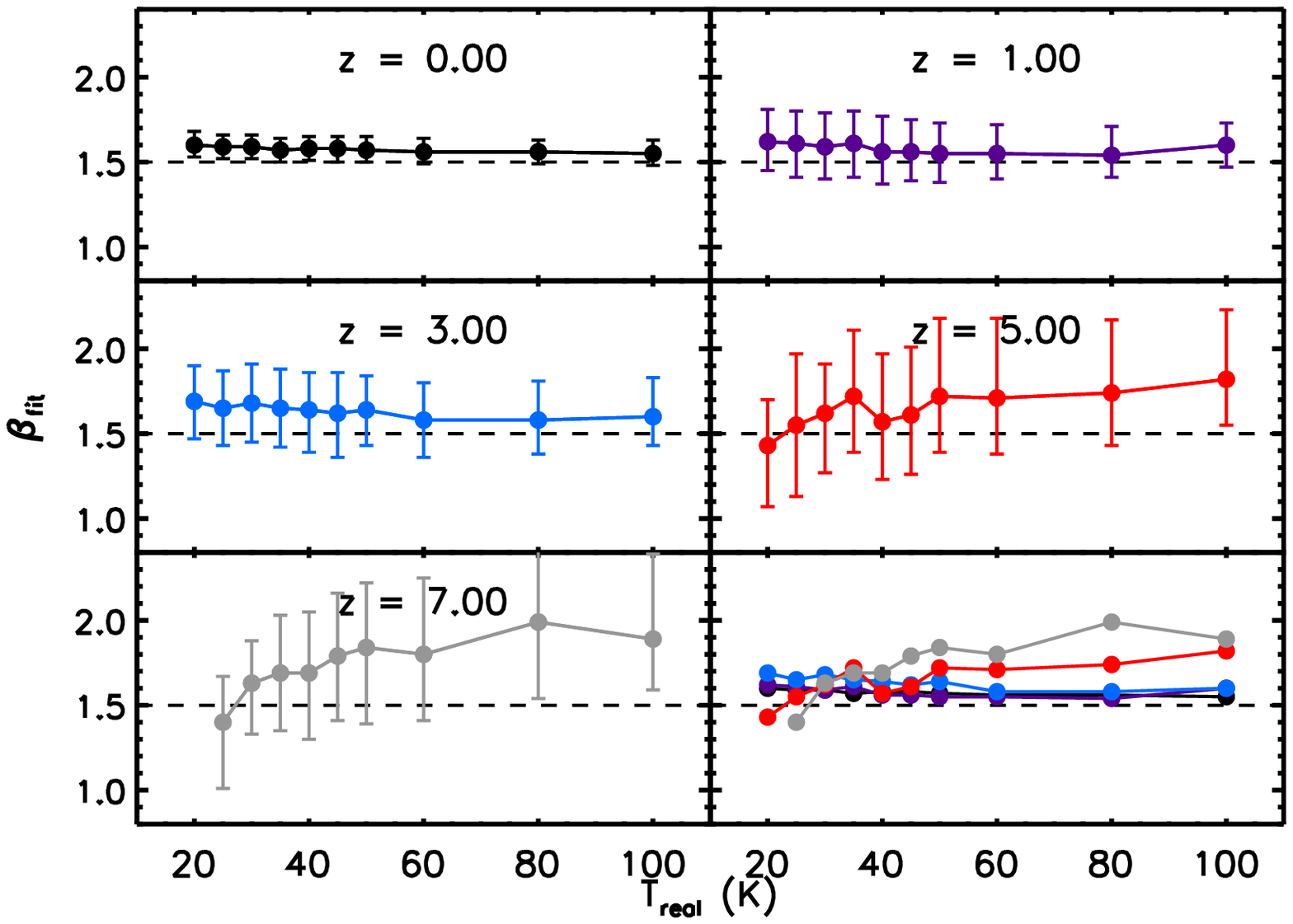}
\caption{
Same as Fig.~\ref{Fig_massfit_MBBtest}, but showing $\beta$ rather than \mfit.
}
\label{Fig_beta_MBBtest}
\end{center}
\end{minipage}
\end{figure*}

\begin{figure*}
\begin{minipage}{\textwidth}
\begin{center}
\includegraphics[width=.9\hsize]{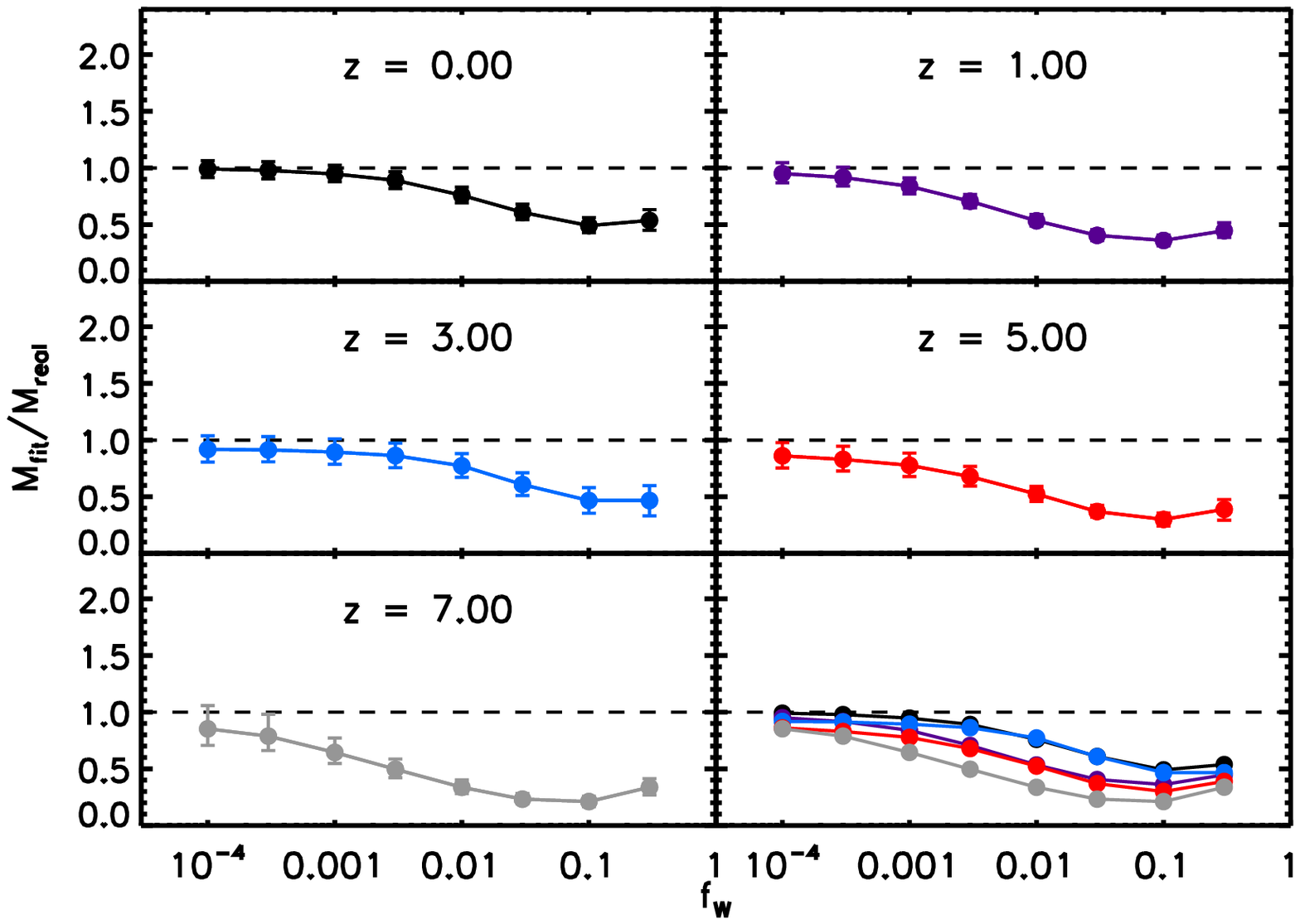}
\caption{
Same as Fig.~\ref{Fig_massfit_MBBtest}, but showing \mfit\ for the two-temperature model. 
}
\label{Fig_massfit_MBBtest2T}
\end{center}
\end{minipage}
\end{figure*}

\begin{figure*}
\begin{minipage}{\textwidth}
\begin{center}
\includegraphics[width=.9\hsize]{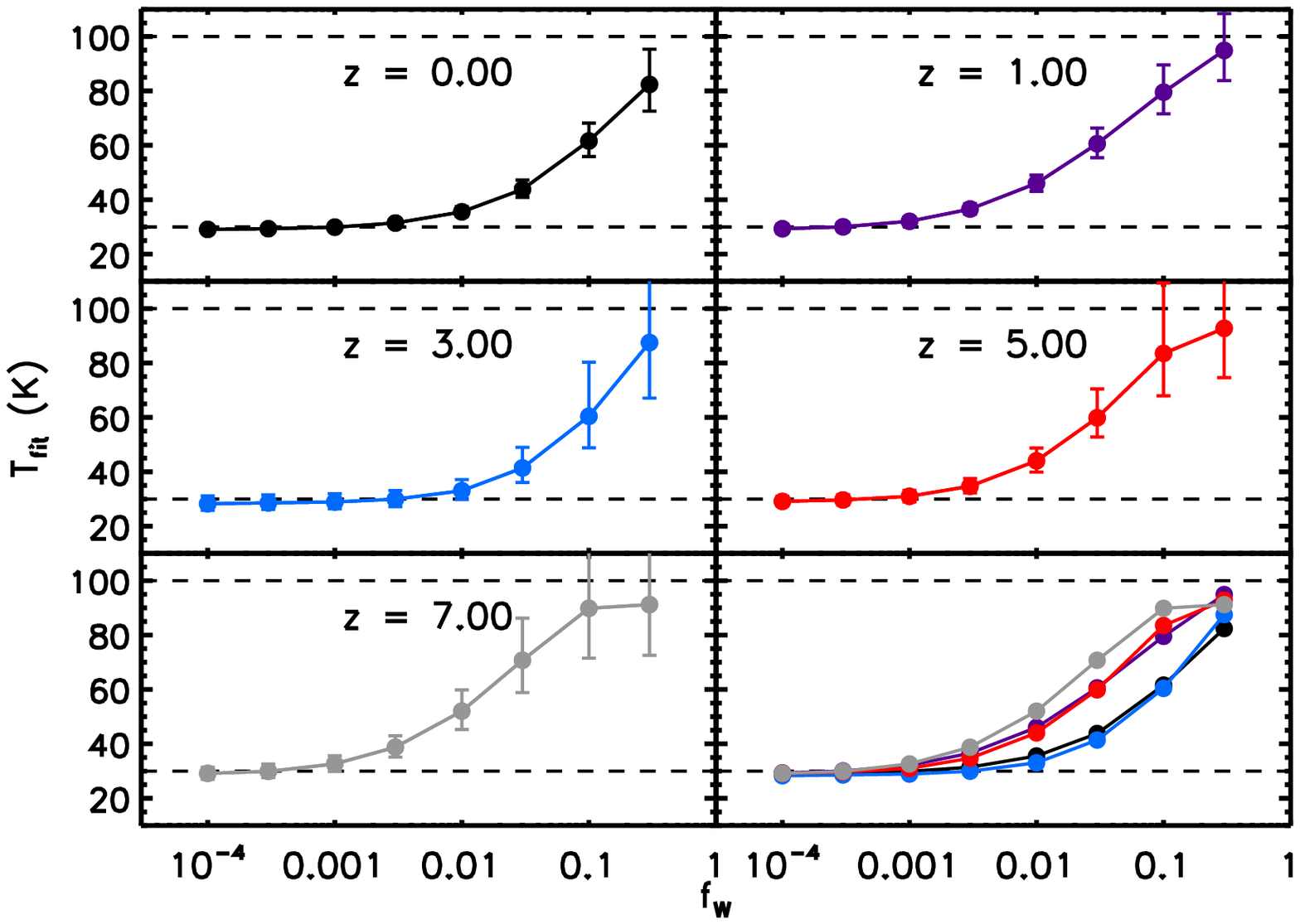}
\caption{
Same as Fig.~\ref{Fig_massfit_MBBtest2T}, but showing \tfit\ rather than \mfit.
}
\label{Fig_Tfit_MBBtest2T}
\end{center}
\end{minipage}
\end{figure*}

\begin{figure*}
\begin{minipage}{\textwidth}
\begin{center}
\includegraphics[width=.9\hsize]{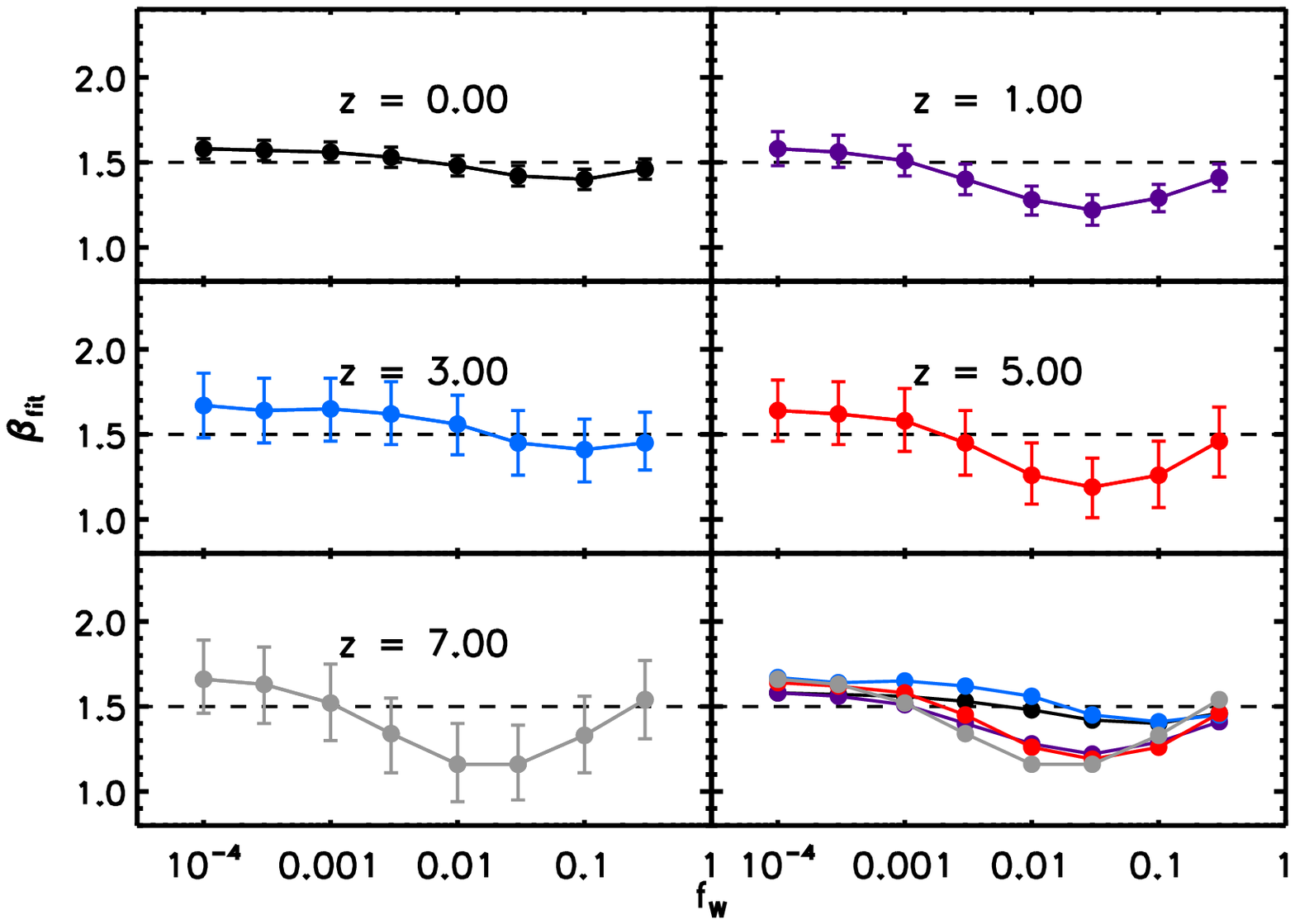}
\caption{
Same as Fig.~\ref{Fig_massfit_MBBtest2T}, but showing $\beta$ rather than \mfit.
}
\label{Fig_beta_MBBtest2T}
\end{center}
\end{minipage}
\end{figure*}

A sanity check for our fitting script is to ensure that it recovers the real values of dust mass and temperature when the optical properties used to make and fit the SED are the same. We created a set of ``test'' SEDs, following the same procedure described in Sect.~\ref{section_model}, but using the opacity from \citet{James+02} -- $\kappa = 0.7\,{\rm cm}^2\,{\rm g}^{-1}$ at $\lambda_0 = 850\,\mu$m. Since $\beta$ is a free parameter in the fit, its value can be chosen arbitrarily for the test model; we adopted a value of 1.5. The modified blackbody is not limited to wavelengths $\leq 1000\,\mu$m, as was the case for \citetalias{Demyk+17A},B data, so the SEDs in this section tend to have more bands (between 4 and 11). The fit results are shown in Figs.~\ref{Fig_massfit_MBBtest} to \ref{Fig_beta_MBBtest2T}; these include the results for single-temperature dust, two-temperature dust, and two-band high-redshift fits.

In the case of single-temperature dust (compare Sect.~\ref{section_fit_dustmass}) the values of \mfit, \tfit\ and \bfit\ from the fit of the test SED correspond to \mreal, \treal\ and $\beta_{\rm real}$ within the uncertainties, as can be seen in Figs.~\ref{Fig_massfit_MBBtest} to \ref{Fig_beta_MBBtest}. The highest redshifts (z = 5 and 7) tend to have less precise results, but they still recover the real parameter values within the error bars. 

The fit results for two-temperature dust (compare Sect.~\ref{section_fit_multiT}) are shown in Fig.~\ref{Fig_massfit_MBBtest2T}, \ref{Fig_Tfit_MBBtest2T} and \ref{Fig_beta_MBBtest2T} for \mfit, \tfit\ and \bfit\ respectively. For small values of \fw, the fit recovers the correct results: \mfit/\mreal\ $\sim$ 1, \tfit\ $\sim$ \treal\ of the cold component, $\beta \sim 1.5$. As \fw\ increases we see both \mfit/\mreal\ and $\beta$ decrease then increase again, in a qualitatively similar way to the fits in Sect.~\ref{section_fit_multiT}, although the shape of the curves so defined -- and especially the depth and location of their minima -- strongly depend on z. This redshift dependence is probably a consequence of the fact that the same bands probe different rest wavelengths at different z.

\begin{figure*}
\begin{minipage}{\textwidth}
\begin{center}
\includegraphics[width=.9\hsize]{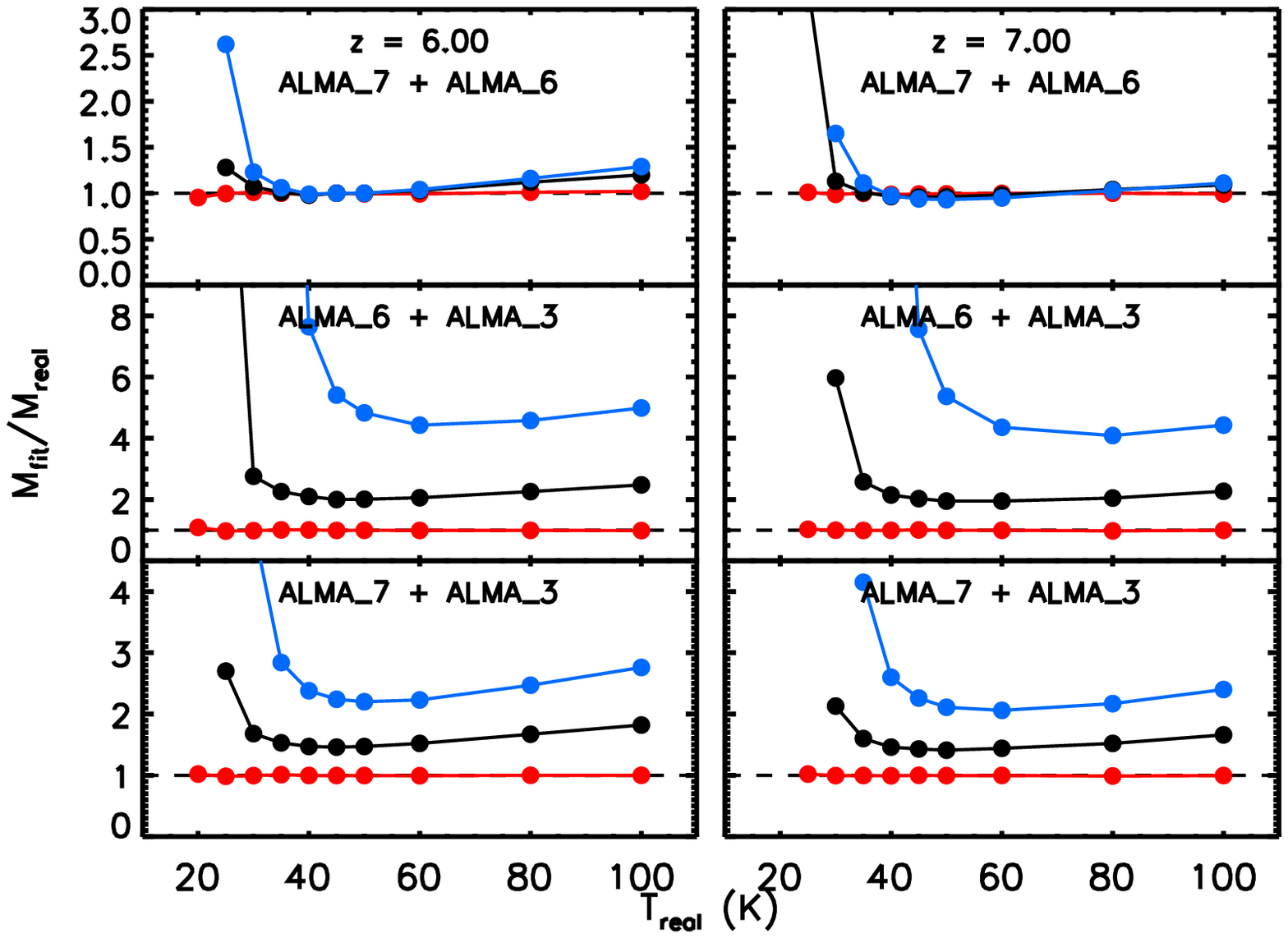}
\caption{
\mfit\ results for high-redshift, two-band fits (compare Fig.~\ref{Fig_2band_Mfit}).
}
\label{Fig_2band_MBBtest_Mfit}
\end{center}
\end{minipage}
\end{figure*}

\begin{figure*}
\begin{minipage}{\textwidth}
\begin{center}
\includegraphics[width=.9\hsize]{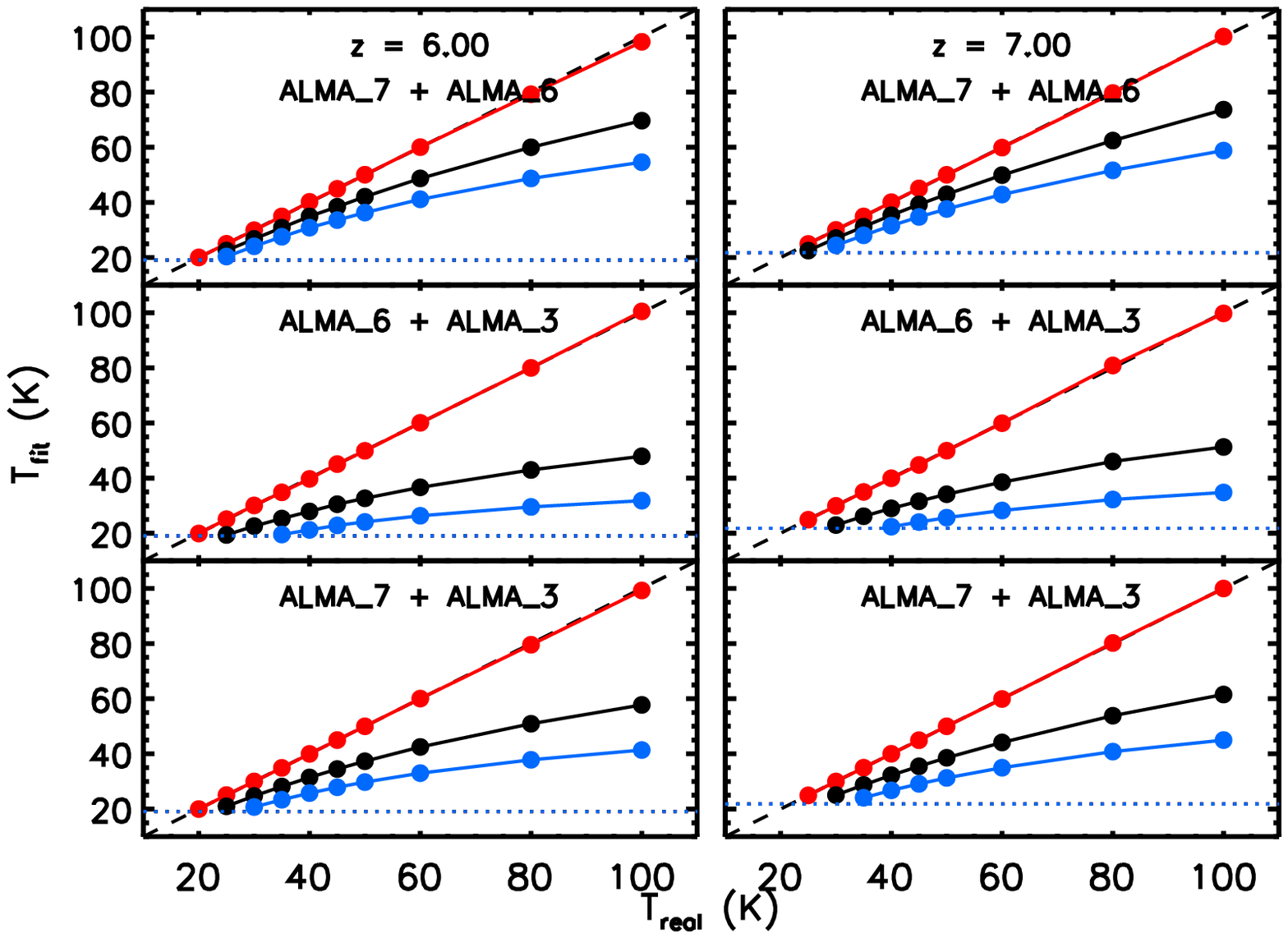}
\caption{
Same as Fig.~\ref{Fig_2band_MBBtest_Mfit}, but for \tfit\ results.
}
\label{Fig_2band_MBBtest_Tfit}
\end{center}
\end{minipage}
\end{figure*}

The fits to the high-redshift, two-band photometry (compare Sect.~\ref{section_fit_2bands}) are shown in Fig.~\ref{Fig_2band_MBBtest_Mfit} (\mfit/\mreal) and \ref{Fig_2band_MBBtest_Tfit} (\tfit). For ALMA bands 6 and 7, the correct dust mass is recovered for all values of $\beta$, with the exception of high-$\beta$, low-\treal\ cases where $C_{\rm CMB}$ is underestimated. For all other band combinations, while the $\beta = 1.5$ fit gives -- unsurprisingly -- perfect results, larger values of $\beta$ result in systematically overestimated dust masses.


\bsp	
\label{lastpage}
\end{document}